\documentclass[book,accept,moreauthors]{Definitions/mdpi}

\providecommand{\authorcontributions}[1]{\par\noindent\textbf{Author Contributions:} #1\par}
\providecommand{\funding}[1]{\par\noindent\textbf{Funding:} #1\par}

\providecommand{\dataavailability}[1]{\par\noindent\textbf{Data Availability Statement:} #1\par}
\providecommand{\conflictsofinterest}[1]{\par\noindent\textbf{Conflicts of Interest:} #1\par}
\providecommand{\appendixtitles}[1]{}
\providecommand{\appendixstart}{}
\providecommand{\reftitle}[1]{} 
\providecommand{\TitleCitation}[1]{}
\providecommand{\AuthorCitation}[1]{}
\providecommand{\hreflink}[1]{}
\providecommand{\PublishersNote}[1]{}

\usepackage{changepage} 
\ifdefined\extralength\else\newlength{\extralength}\fi
\usepackage[normalem]{ulem} 

\newcommand{\ie}{i.e.}
\newcommand{\eg}{e.g.}
\newcommand{\rmd}{\ensuremath{\mathrm{d}}}

\newcommand{\za}{\ensuremath{\alpha}}   
\newcommand{\zb}{\ensuremath{\beta}}    
\newcommand{\zg}{\ensuremath{\nu}}      
\newcommand{\moment}{\ensuremath{q}}
\newcommand{\qmom}{\ensuremath{q^{\text{th}}}}

\providecommand{\eq}{}\renewcommand{\eq}[1]{(\ref{eq:#1})}
\providecommand{\eqn}{}\renewcommand{\eqn}[1]{Equation~\eq{#1}}
\providecommand{\eqns}{}\renewcommand{\eqns}[1]{Equations~\eq{#1}}
\providecommand{\fignum}{}\renewcommand{\fignum}[1]{\ref{fig:#1}}
\providecommand{\figref}{}\renewcommand{\figref}[1]{Figure~\ref{fig:#1}}
\providecommand{\figrefs}{}\renewcommand{\figrefs}[1]{Figures~\ref{fig:#1}}
\providecommand{\secref}{}\renewcommand{\secref}[1]{Section~\ref{sec:#1}}
\providecommand{\appref}{}\renewcommand{\appref}[1]{Appendix~\ref{app:#1}}
\providecommand{\tabref}{}\renewcommand{\tabref}[1]{Table~\ref{tab:#1}}

\firstpage{1}

\pubvolume{1}
\issuenum{1}
\articlenumber{0}
\pubyear{2026}
\copyrightyear{2026}
\datereceived{ }
\daterevised{ }
\dateaccepted{ }
\datepublished{ }
\hreflink{https://doi.org/}

\Title{Universality in Anomalous Diffusion: Power Laws and Faster-Decaying Tails}
\TitleCitation{Universality in Strong Anomalous Diffusion: Power Laws and Faster-Decaying Tails}

\Author{Claudio Giberti $^{1}$ and Alessandro Vezzani$^{4,5,6}$ and Raffaella Burioni$^{4,5}$ and Lamberto Rondoni $^{2,3,*}$}
\AuthorNames{Claudio Giberti and Alessandro Vezzani and Raffaella Burioni and Lamberto Rondoni}
\AuthorCitation{Giberti, C.; Vezzani, A.; Burioni, R.; Rondoni, L.}

\abstract{Strong anomalous diffusion is often characterized by a piecewise-linear spectrum of the moments of displacement.
The spectrum is characterized by the slopes $\theta$ and $\zeta$ for small and large moments, respectively, and by the crossover
moment $q_c$ at which the slope changes. The exponents $\theta$ and $\zeta$ characterize the asymptotic scaling of the bulk and of the
tails of the probability distribution function of displacements, respectively. Here, we adopt asymptotic theory to match the
behaviors at intermediate scales. The resulting constraint explains how distributions with algebraic tails imply strong
anomalous diffusion, and it relates $q_c$ to the corresponding power law. Our theory provides universal hyper-scaling
relations between the exponents characterizing strong anomalous diffusion, and it yields explicit expressions for the
leading-order corrections to the asymptotic power-law behavior of the moments of displacement. 
We also characterize finite-time corrections and propose a robust distribution-based method to estimate $\theta$, $\zeta$, and $q_c$.
The findings are supported by numerical and analytical results for three paradigmatic models
exhibiting strong anomalous diffusion.
Finally, we show that power-law tails are not only sufficient but also necessary for
this scenario: 
with a Gaussian, or generally fast-decaying, bulk and exponentially rare ballistic
excursions, the spectrum of moments is single-linear, and the crossover to the apparent second branch drifts
with the observation time. When the bulk is fast decaying and the tail is a power law, strong anomalous
diffusion is still present, but the hyper-scaling relations among the exponents are violated.}


\begin{document}

\renewcommand{\thefootnote}{\fnsymbol{footnote}}

\footnote[0]{%
$^{1}$ \quad Dipartimento di Scienze e Metodi dell'Ingegneria, Universit\`a di
Modena e Reggio Emilia, Via Amendola 2, Padiglione Morselli, I-42100 Reggio Emilia, Italy\\
$^{2}$ \quad Department of Mathematical Sciences, Politecnico di Torino, Corso Duca degli Abruzzi 24, I-10129 Torino, Italy\\
$^{3}$ \quad INFN, Sezione di Torino, Via P.\ Giuria 1, 10125 Torino, Italy\\
$^{4}$ \quad Department of Mathematical, Physical and Computer Science, Universit\`a degli Studi di Parma, Parco Area delle Scienze, 7/A 43124 Parma, Italy\\
$^{5}$ \quad INFN, Sezione di Milano Bicocca, Piazza della Scienza 3, 20126, Milano, Italy\\
$^{6}$ \quad Istituto dei Materiali per l'Elettronica ed il Magnetismo (IMEM-CNR), Parco Area delle Scienze, 37/A, 43124 Parma, Italy\\
$^*${Correspondence: lamberto.rondoni@polito.it}
}
\section{Introduction}
\label{sec:intro}


Great theoretical efforts have been devoted to unveiling the physical mechanisms that induce
anomalous diffusion (see, \eg, Refs.~\cite{Metzler,KRS08,Sok12} and references therein) in systems where the stochastic motion cannot be described by standard Brownian motion. They have been accompanied by a
rapid increase of highly resolved experimental observations, particularly of video microscopy and particle tracking, as well
as by ever more refined numerical simulations of ensembles of trajectories \cite{frontiers}. Processes exhibiting anomalous
diffusion are observed in a broad range of phenomena, including molecules moving in the living cell \cite{Saxton,GC06},
dynamics on cell membranes \cite{NHB07}, crowded environments \cite{Sok12}, soil transport \cite{Rina}, heat transport in
low-dimensional systems \cite{LLP}, and certain classes of billiards \cite{AHO03,JBR08}, to mention but a few.

In these processes the probability distribution function (PDF) of the displacement is commonly non-Gaussian, meaning that the
signatures of anomalous diffusion appear not only in the mean square displacement (MSD), but essentially in all the moments of the distribution. Deviations from
Gaussian behavior are observed in the tails and, in certain dynamics, the large fluctuations of the tails scale differently
from the small fluctuations described by the bulk of the distribution. In general, denoting by $r(t) := |x(t) - x(0)|$ the
displacement accumulated over a time $t$---a random variable whose generic value we denote by $r$---the time
scaling of its \qmom{} moment can be written as
\begin{subequations} \label{eq:spectrum}
\begin{align} \label{eq:spectrum-evolution}
  \left\langle r^\moment \right\rangle_t
  \sim t^{\moment\, \zg_\moment /2} \, ,
\end{align}
where $\zg_\moment = 1$ indicates normal diffusion, as derived from Fick's law, and $\zg_\moment \neq 1$ anomalous diffusion.
A scale-invariant transport with a constant $\zg_\moment \neq 1$  is called (weakly) anomalous. 
While, if $\zg_\moment$ varies with the moment order $\moment$,
scale invariance breaks down, and transport is said to be strongly anomalous \cite{CMMGA99,VRTGM21}. 
As already noticed in \cite{CMMGA99} there are processes where $\zg_{\moment=2} = 1$ but $\zg_\moment$ depends on the moment $\moment$;  therefore a strong anomalous behavior can be observed even if   $\left\langle r^2 \right\rangle  \sim t$ and the MSD seems to suggests a regular diffusion.

The most commonly
investigated strong anomalous situation is characterized by two regimes, in which the power of $t$ is given by two different
expressions that are linear in the moment order:
\begin{align} \label{eq:momentPiecewise}
  \frac{\moment\, \zg_\moment}{2} = \left\{
  \begin{array}{lll}
    \theta \, \moment                            & \text{ for} & \moment \leq q_c \, ,
    \\[1mm]
    \zeta \, \moment - q_c \, (\zeta - \theta )  & \text{ for} & \moment > q_c \, ,
  \end{array}
  \right.
\end{align}%
\end{subequations}%
where $\theta$, $\zeta$, and $q_c$ are given parameters. Indeed, strong anomalous diffusion repeatedly arises as the combination
of anomalous diffusion determining a local scale invariance for the small fluctuations of the displacement (with
$\moment \leq q_c$), and rare but long-lasting ballistic excursions determining a faster scaling of the large fluctuations,
for which $\zeta = 1$. The piecewise-linear behavior of \eqn{momentPiecewise} has been observed in the transport in polygonal
billiards \cite{SL06,JR06,Orchard21}, billiards with infinite horizon \cite{AHO03,SS06,CESZ08}, diffusion in laser-cooled
atoms \cite{AKB17}, one-dimensional maps \cite{Pik91,SRGK15}, intermittent maps \cite{CMMGA99,AC03}, experiments on the
mobility of particles inside living cancer cells \cite{GW10}, the motion of particles passively advected by dynamical
membranes \cite{SJ94}, running sand piles \cite{CLNZ99}, and stochastic models of inhomogeneous media \cite{BCV10,BVLV14},
among a host of other phenomena. A great deal of insight into strong anomalous diffusion has been gained by means of
generalizations of the central limit theorem in terms of non-normalizable densities \cite{RDHB14a,RDHB14b,FSBZ15,ZDK15}
and the big jump principle \cite{VBB19,VBB20,ENBVB26}. The
above scenario is not unique---for instance, in the bulk-mediated diffusion on lipid bilayers of Ref.~\cite{KCNP16} the
spectrum of moments is not piecewise linear---but the double linear scaling of \eqn{momentPiecewise} is the most commonly
reported situation in experimental observations and theoretical studies, and it is the one investigated here, for a set of
different deterministic and stochastic systems.

We revisit the theory of strong anomalous diffusion from the point of view of scaling theory \cite{Barenblatt} and asymptotic
matching \cite{BO} of the PDF $\mathcal{P}(r, t)$ of the displacement, developing the approach of Ref.~\cite{arxiv}, which is
summarized below. In that paper, the process $R(t)$, which takes values $r$, was assumed to evolve on two different asymptotic scales, described by
two rescaled versions of the displacement, $z := r/t^\theta$ and $y := r/t^\zeta$, which pertain to two different regions of
space: for sufficiently large times $t$, the PDF is subdivided into two branches, corresponding to two distinct asymptotic
regimes, which coincide over an interval that gets larger and larger as $t$ grows. This was formalized through the following
\emph{ansatz}:
\begin{align}\label{eq:ProbScalingCutoff}
  \mathcal{P}(r, t) \, \rmd r
  = \left\{
  \begin{array}{rll}
    \mathcal{F}(z) \, \rmd z
    & \text{with}
    & z = \frac{r}{t^\theta} \leq c\, t^{\iota-\theta} \, ,
    \\[2mm]
    g(t) \, \mathcal{J}(y) \, \rmd y
    & \text{with}
    & y = \frac{r}{t^\zeta} > c \, t^{\iota-\zeta} \, ,
  \end{array}
  \right.
\end{align}
where $\theta$, $\iota$, $\zeta$, and $c$ are four parameters obeying $0<\theta<\iota<\zeta$ and $c>0$, and the scaled variables
$z$ and $y$ concern the ``bulk'' and the ``tail'' of $\mathcal P$, respectively defined by $r \in [0, c\, t^\iota)$ and
$r \in (c\, t^\iota, \infty)$. The functions $\mathcal{F}$ and $g\,\mathcal{J}$ are non-negative and integrable on their
respective domains, so that $\mathcal P$ is non-negative and normalized over the half real line, for all $t$. The domain of
$\mathcal{F}(z)$ grows without bounds with $t$, since the value of $z$ at its right border, $c\, t^{\iota-\theta}$, approaches
infinity in the large-time limit. Analogously, the domain of $\mathcal{J}(y)$ extends to arbitrarily small values, since the
smallest admissible value $c\, t^{\iota-\zeta}$ of $y$ approaches $0$ for $t \to \infty$. Throughout the paper, only these
three variables are needed: $r$ is the physical displacement, while $z$ and $y$ measure it in units of the bulk scale $t^\theta$
and of the tail scale $t^\zeta$, respectively.

\begin{figure}[H]
  \includegraphics[width=0.48\linewidth]{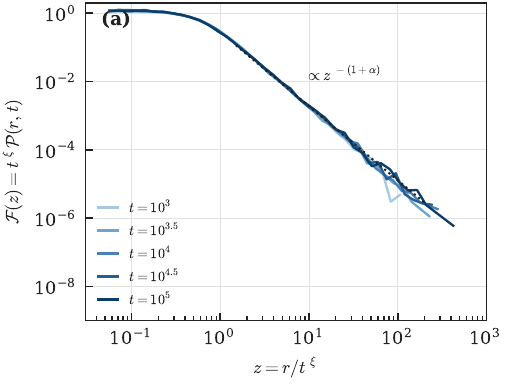}\hfill
  \includegraphics[width=0.48\linewidth]{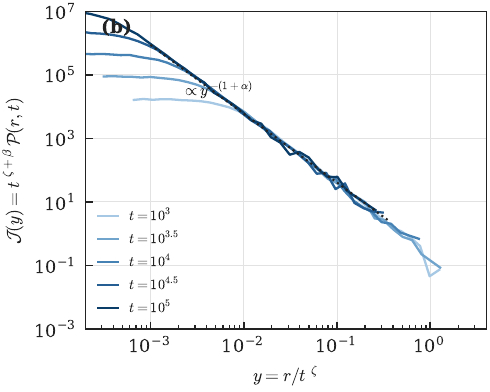}
  \caption{Scaling of (\textbf{a}) the bulk and (\textbf{b}) the tails of the PDF for the SDCA with $D = 0.4$, \ie{}
  $\za = p_6 = 7/6$. The color of the lines refers to the times $t$ provided in the legends. The dotted lines show the
  asymptotic power-law scaling for large $z$ and small $y$, respectively; the power law has the exponent $-(1+\za) = -13/6$,
  while $\theta = 1/p_6 = 6/7$ and $\zeta = 3/2$. Details on the model and on the numerical parameter values are provided in
  \secref{SDCA}.}
  \label{fig:PDFscaling}
\end{figure}

Here, we illustrate first the validity of this hypothesis for the Spatial Diffusion of Cold Atoms (SDCA) in shallow optical
lattices, an experimentally realized system of anomalous transport that is described in \secref{SDCA}. \figref{PDFscaling}
shows how, in this case, the scaling function $\mathcal{F}(z)$ takes the form of a power law for $z > z_c \simeq 2$, and
$\mathcal{J}(y)$ follows a power law for $y < y_c \simeq 0.25$. Reference~\cite{arxiv} then used the fact that the two
branches of $\mathcal{P}(r, t)$ overlap as a power law in the interval $z_c\, t^\theta \lesssim r \lesssim y_c\, t^\zeta$, and
that this interval expands for increasing $t$. The PDF thus enjoys asymptotic matching: as long as $\theta < \iota < \zeta$ and
$t$ is large, the matching position of the two branches, $r = c\, t^\iota$, resides well inside this interval, irrespective
of the particular choice of $c$ and $\iota$.

The present paper builds on these results, as well as on those summarized in the next sections and in the appendices: the
hyper-scaling relation \eq{betadef} with the ensuing piecewise-linear spectrum; the corrections to scaling of
\appref{fitProblems}; and the fitting scheme of \appref{parameterFitting}. These were tested in Ref.~\cite{arxiv} on quenched
disordered systems of the L\'evy--Lorentz type, on Lorentz gases, and on polygonal billiards. Here we report new numerical
results on the SDCA---an experimentally realized dynamics whose non-ballistic cutoff exponent puts the full three-exponent
fitting scheme to work---together with the finite-sampling criterion of \secref{SDCA}; and the entire converse theory of
\secref{gaussian},  which establishes that power-law tails, 
carried by a weight $g(t)$ that itself decays as a power of
time, are not only sufficient but also necessary for strong anomalous
diffusion. 
In particular for exponentially rare
ballistic flights our analysis characterizes a phenomenon that we call transient imitation. 

In \secref{moments} we show how the piecewise-linear scaling of $\zg_\moment$ emerges as a consequence of
\eqn{ProbScalingCutoff}. Moreover, an asymptotic matching of the bulk and the tail of $\mathcal{P}(r,t)$ can be imposed 
in the presence of power-law decay of $\mathcal{F}(z)$, $\mathcal{J}(y)$ and $g(t)$.
This matching establishes universal hyper-scaling relations between
the exponents characterizing the system. In particular the crossover moment $q_c$ equals the exponent $\alpha$ that describes the decay of the scaling functions. In \secref{discussion} we
explore these findings for three models that feature anomalous transport: besides elaborating on the SDCA, we revisit the
exact solution of the Slicer Map (SM) \cite{SRGK15,VRTGM21}, and we verify the theory on numerical data for L\'evy walks (LW)
\cite{Metzler,ZDK15}. The distributions and the spectra of moments of all these systems are described by the theory
exemplified in \secref{moments}.  In \secref{gaussian} we then present the converse theory: we prove that asymptotic
matching itself forces 
the intermediate asymptotics---the large-$z$ tail
of $\mathcal{F}(z)$--- to be a power law. 
In particular if the bulk $\mathcal{F}(z)$ is Gaussian and the excursions are exponentially rare,
the apparent crossover of moments drifts as the observation time grows and
the moment spectrum is single-linear.
For a Gaussian bulk coexisting with a power-law tail, instead, asymptotic matching cannot be realized:
strong anomalous diffusion persists, with a spectrum of the form of \eqn{momentPiecewise}, but the hyper-scaling
relation between the crossover moment and the tail exponent is violated. \secref{solvable} is devoted to exactly solvable deterministic models, within the Slicer Map class and
its variants, that realize all these scenarios.
In \secref{conclusion} we summarize our main findings
and provide an outlook on future perspectives. \appref{fastCutoff} contains the proof that no strong anomalous diffusion can
arise when the tails of the distribution decay faster than any power law.
\section{Scaling and Asymptotic Matching}
\label{sec:moments}

In this section we summarize the results of Ref.~\cite{arxiv}, whose novel application to the SDCA is presented in
\secref{SDCA}. We begin by assuming that at large times $t$ the time evolution of the bulk of the probability density obeys
scaling with the scaling variable $z = r / t^\theta$ (\eqn{ProbScalingCutoff}), and that it assumes a power law for large
displacements,
\begin{subequations}\label{eq:match}
\begin{align}
\label{eq:ProbTailX}
  \mathcal{P}(r, t) \simeq \za \, k \, r^{-\za-1} \, t^{\za\,\theta} \, .
\end{align}

\subsection{Asymptotic Matching and Hyper-Scaling}
\label{sec:matching}

Given the definition \eq{ProbScalingCutoff}, the power law \eq{ProbTailX} may be written for the two branches of the PDF. The
bulk is described by the scaling function
\begin{align}
  \mathcal{F}(z)
  \simeq \za\, k \, r^{-\za-1} \, t^{\za\,\theta} \, \frac{\rmd r}{\rmd z}
  = \za\, k \, z^{-\za-1}
  \qquad \text{for } z > z_c \, ,
  \label{eq:ProbTailZ}
\end{align}
where the parameter $z_c$ is deduced from the data. For the tail, we rely on the function $g$ introduced in
\eqn{ProbScalingCutoff}, noting that \eqn{ProbTailX} implies
\begin{align} \label{eq:ProbV}
  \mathcal{J}(y)
  \simeq
  \frac{\za k \, r^{-\za-1}}{g(t)} \, t^{\za \theta} \, \frac{\rmd r}{\rmd y}
  = \frac{\za k \, r^{-\za-1}}{g(t)} \, t^{\za\,\theta + \zeta} \, .
\end{align}
Then, taking
\begin{align} \label{eq:betadef}
  g(t) = t^{-\zb} \quad \text{ with } \quad \zb = \za \, (\zeta-\theta) \, ,
\end{align}
one obtains
\begin{align}
  \mathcal{J}(y)
  = \za\, k \, y^{-\za-1}
  \qquad \text{for } y < y_c \, ,
\label{eq:ProbTailV}
\end{align}
\end{subequations}
where $y_c$ is deduced from the data, like $z_c$ is. At late times the bulk and the tail of the distribution overlap as power
laws in a wide interval whose width diverges asymptotically in time.

There are two important things to note here. Firstly, in terms of the scaled variables and distributions, the deviation from
the power law can be pushed below any desired bound by selecting larger values of $z_c$ and smaller values of $y_c$. Even
when $y_c \ll z_c$, it suffices to wait for large values of $t$ until the interval $z_c\, t^\theta < r < y_c\, t^\zeta$ is
non-empty. Eventually, in the large-$t$ limit, the length of the interval diverges because $\theta < \zeta$.
Secondly, in \eqns{ProbTailZ} and \eq{ProbTailV} not only the exponent $-(\za+1)$ of the power laws, but also the prefactors,
$\za k$, agree. This holds for every set of exponents $\{\za, \zb, \theta, \zeta\}$ characterizing the PDF. The constraint
\eq{betadef} is a hyper-scaling relation that expresses the exponent $\zb$, governing the temporal decay of the tail
amplitude, in terms of $\za$, $\zeta$, and $\theta$; the agreement of the prefactors will become the cornerstone of the data
analysis presented in \appref{parameterFitting}.

\medskip
One may wonder whether some non-algebraic shape---for instance, a Gaussian---could play the role of the
\emph{intermediate} asymptotics joining the bulk and the tail branches, with the power law \eq{ProbTailX} suitably replaced.
The answer is negative: the matching structure by itself selects power laws. 

\begin{Proposition}\label{prop:rigidity}
  Let the ansatz \eq{ProbScalingCutoff} hold with $0 < \theta < \iota < \zeta$, and assume that the two branches simultaneously
  describe $\mathcal P$ on the overlap region: there are positive, continuously differentiable functions $\varphi$ (the
  large-$z$ form of $\mathcal F$) and $\psi$ (the small-$y$ form of $\mathcal J$), and a positive function $g$, such that
  \begin{align}\label{eq:matchFunctional}
    t^{-\theta} \, \varphi\!\left( r \, t^{-\theta} \right)
    = g(t) \; t^{-\zeta} \, \psi\!\left( r \, t^{-\zeta} \right)
  \end{align}
  for all $r \in (z_c\, t^\theta ,\, y_c\, t^\zeta)$ and all 
  $t>t_c$,  where $z_c,\, y_c,\, t_c$ are given positive constants. Then there exist constants $\za > 0$ and
  $k > 0$ such that
  \begin{align*}
    \varphi(z) = \za\, k \, z^{-\za-1} \, ,
    \qquad
    \psi(y) = \za\, k \, y^{-\za-1} \, ,
    \qquad
    g(t) = t^{-\za\,(\zeta-\theta)} \, .
  \end{align*}
\end{Proposition}

\begin{proof}
  Taking the logarithm of \eqn{matchFunctional} and differentiating with respect to $\ln r$ at fixed $t$ gives
  $\Lambda_\varphi(z) = \Lambda_\psi(y)$, where $\Lambda_f(u) := u\, f'(u)/f(u)$ is the logarithmic slope,
  $z = r\, t^{-\theta}$, and $y = r\, t^{-\zeta}$  are the scaled variable in the "bulk" and the "tail" of $\mathcal P$. Fix $\bar z > z_c$ and set $r = \bar z\, t^{\theta}$: this point remains in the
  overlap for all sufficiently large $t$, while $y = \bar z\, t^{\theta-\zeta}$ sweeps, continuously in $t$, an interval
  accumulating at $0$. 
  Therefore, since every point in that interval can be written as $y = \bar z\, t^{\theta-\zeta}$, we have $\Lambda_\psi(y) = \Lambda_{\varphi}(\bar z)$. 
  Hence $\Lambda_\psi(y)$ is constant, 
  on that interval; and since
  $\bar z$ is arbitrary, $\Lambda_\varphi (z)$ is constant on $(z_c, \infty)$ as well. Call the common value of the logarithmic slopes $-(\za+1)$.
  Integrating $\Lambda_f(u)=-(\alpha+1)$, we obtain $\varphi(z) = k_1\, z^{-\za-1}$ and $\psi(y) = k_2\, y^{-\za-1}$; integrability of $\mathcal F$ at large $z$,
  required by the normalization of $\mathcal P$, imposes $\za > 0$. Substituting back into \eqn{matchFunctional} yields
  $k_1\, t^{\za\theta} = g(t)\, k_2\, t^{\za\zeta}$, whence $g(t) = (k_1/k_2)\, t^{-\za(\zeta-\theta)}$. Normalizing $g$ as in
  \eqn{betadef} absorbs the constant $k_1/k_2$, and gives $k_1 = k_2 = \za k$ together with the hyper-scaling relation
  $\zb = \za\,(\zeta-\theta)$.
\end{proof}

\begin{Corollary}\label{cor:noGauss}
  If the large-$z$ asymptotics of $\mathcal F$ decays faster than any power law, then
  no pair $(\psi, g)$ satisfies \eqn{matchFunctional} on an asymptotically widening overlap. Within the
ansatz \eq{ProbScalingCutoff}, the tail branch is then either absent, yielding a single linear spectrum, 
  or it is separated from the bulk by a region in which $\mathcal P$ is
  negligible.
\end{Corollary}

Proposition~\ref{prop:rigidity} shows that the power law \eq{ProbTailX} is not an assumption of convenience: it is the
\emph{only} intermediate asymptotics compatible with the matching of two different self-similar branches, in the spirit of
intermediate-asymptotics theory \cite{Barenblatt}. In this sense, power-law tails are necessary
and, as it will be shown in the next subsection, sufficient --- for a time-persistent, bi-linear spectrum of moments. 
That the absence of the tail branch, considered in the first alternative of the previous Corollary, leads to the single linear spectrum follows from the analysis carried out in \appref{fastCutoff}. An example with a Gaussian bulk is provided in Section 4.1.


\subsection{Moments}
\label{sec:momentScalings}

The \qmom{} moment of the displacement is defined by
\begin{align*}
  \left\langle r^\moment \right\rangle_t
  := \int_{0}^\infty r^\moment \, \mathcal{P}(r, t) \, \rmd r \, .
\end{align*}
For large $t$, where $z_c\, t^\theta < c\, t^\iota < y_c\, t^\zeta$, the integral can be broken into three pieces: the bulk
contribution up to $z_c\, t^\theta$, the intermediate power-law region, and the tail contribution beyond $y_c\, t^\zeta$. In
view of \eqn{ProbScalingCutoff}, in which the Jacobians of the coordinate transformations are already accounted for, and
applying \eqn{ProbTailX} in the intermediate region, the \qmom{} moment can be approximated asymptotically in time as
\begin{align}\label{eq:momentSplitting}
  \left\langle r^\moment \right\rangle_t
  \simeq t^{\moment\,\theta} \int_{0}^{z_c} z^\moment \, \mathcal{F}(z) \, \rmd z
  + t^{\moment\,\zeta - \zb} \int_{y_c}^{\infty} y^\moment \, \mathcal{J}(y) \, \rmd y
  + \za \, k \, t^{\za \theta} \int_{z_c\,t^\theta}^{y_c\,t^\zeta} r^{\moment-\za-1} \, \rmd r \, .
\end{align}
The first two integrals take fixed, finite values that do not depend on time. Moreover, for $\moment \neq \za$ the last
integral in \eqn{momentSplitting} can be expressed as the difference of two power laws with the same exponents as those
appearing in front of the other two integrals. Collecting terms with the same time dependence provides
\begin{subequations} \label{eq:momentsQ}
\begin{align} \label{eq:momentQ}
  \left\langle r^\moment \right\rangle_t
  = N_\moment \, t^{\moment\,\theta} + I_\moment \, t^{\moment\,\zeta-\zb}
  \qquad \text{for } \moment \neq \za \, ,
\end{align}
with
\begin{align}
  N_\moment
  &= \frac{\za \, k \, z_c^{\moment-\za}}{\za - \moment}
  + \int_{0}^{z_c} z^\moment \, \mathcal{F}(z) \, \rmd z \, ,
  \label{eq:Nq}
  \\[2mm]
  I_\moment
  &= \frac{\za \, k \, y_c^{\moment-\za}}{\moment-\za}
  + \int_{y_c}^{\infty} y^\moment \, \mathcal{J}(y) \, \rmd y \, .
  \label{eq:Iq}
\end{align}
\end{subequations}
For $\moment = \za$, on the other hand, the third integral in \eqn{momentSplitting} takes the form of a logarithm, and one
finds
\begin{subequations} \label{eq:momentAlpha}
\begin{align}
  \frac{\left\langle r^\za \right\rangle_t}{t^{\za\,\theta}}
  = N_\za + I_\za + \za\, k \, (\zeta-\theta) \, \ln t \, ,
\end{align}
where
\begin{align}
  N_\za &= -\za \, k \, \ln z_c + \int_{0}^{z_c} z^\za \, \mathcal{F}(z) \, \rmd z \, ,
  \\
  I_\za &= \za \, k \, \ln y_c + \int_{y_c}^{\infty} y^\za \, \mathcal{J}(y) \, \rmd y \, .
\end{align}
\end{subequations}
In either case the coefficients $N_\moment$ and $I_\moment$ take finite values as long as $\langle r^\moment \rangle_t$
exists, which is always the case at finite $t$ when the speed of the moving objects is bounded, so that the distribution has
a cutoff. Moreover, in the scaling regime these values do not depend on the particular choice of $z_c$ and $y_c$, as one
verifies by direct computation using \eqns{ProbTailZ} and \eq{ProbTailV}.

It is useful to note the geometric interpretation of the coefficients. Normalization of the PDF implies that $\mathcal F$,
and hence $z^\moment \mathcal F$ for all $\moment>0$, is integrable at small $z$. For $\moment < \za$ the power-law tail is
integrable at infinity, and $N_\moment$ amounts to the integral of $z^\moment \mathcal{F}(z)$ over $\mathbb R_+$; it takes a
positive value. For $\moment > \za$, instead, $N_\moment$ amounts to the integral of the difference
$z^\moment \mathcal{F}(z) - \za k\, z^{\moment-\za-1}$, and it can take positive or negative values, depending on whether
$\mathcal F$ lies above or below the power law. Similarly, $I_\moment$ is positive for $\moment > \za$, where it amounts to
the integral of $y^\moment \mathcal{J}(y)$ over $\mathbb R_+$, while for $\moment < \za$ it is the integral of
$y^\moment \mathcal{J}(y) - \za k\, y^{\moment-\za-1}$, whose sign signifies which portion of $\mathcal{J}$ lies below or
above the power law.

There are three scenarios for the asymptotic scaling of the moments. For $\moment < \za$ one has
$\moment\zeta - \zb = \moment\zeta - \za(\zeta-\theta) < \moment\theta$, so that at large $t$ the right-hand side of \eqn{momentQ}
is dominated by the term containing $N_\moment$. Conversely, for $\moment > \za$ the exponent $\moment\zeta - \zb$ is the
larger one, and the term containing $I_\moment$ dominates. For $\moment = \za$ the logarithmic term dominates in
\eqn{momentAlpha}. Asymptotically in time we may thus write
\begin{align} \label{eq:momentsScaling}
  \left\langle r^\moment \right\rangle_t
  \simeq \left\{
  \begin{array}{lrl}
    N_\moment \; t^{\moment\,\theta} & \quad\text{for} & 0 < \moment < \za \, ,
    \\[2mm]
    \za \, k \, (\zeta-\theta) \; t^{\za\,\theta} \, \ln t & \quad\text{for} & \moment = \za \, ,
    \\[2mm]
    I_\moment \; t^{\moment\,\zeta - \zb} & \quad\text{for} & \za < \moment \, .
  \end{array}
  \right.
\end{align}
In particular, we recover the piecewise-linear spectrum \eq{momentPiecewise}; moreover, a direct
consequence of the asymptotic matching (i.e., of the hyper-scaling relation \eq{betadef}) is that the crossover
moment $q_c$ equals $\alpha$.
The result is purely based on the constraints imposed by matching the power laws of the bulk and of the
tail of the distribution. In \secref{discussion} we present numerical and analytical results confirming that this theoretical
framework applies to a vast variety of cases.

\subsection{Comparison with Numerical Data}
\label{sec:checkMoments}

\begin{figure}[H]
  \includegraphics[width=0.48\linewidth]{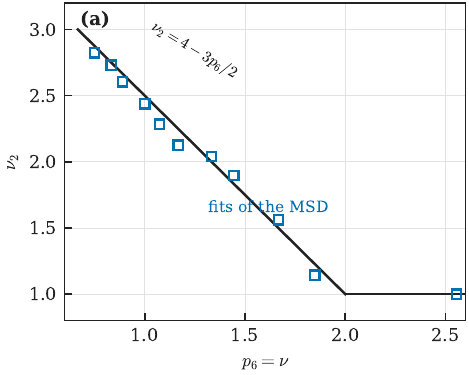}\hfill
  \includegraphics[width=0.48\linewidth]{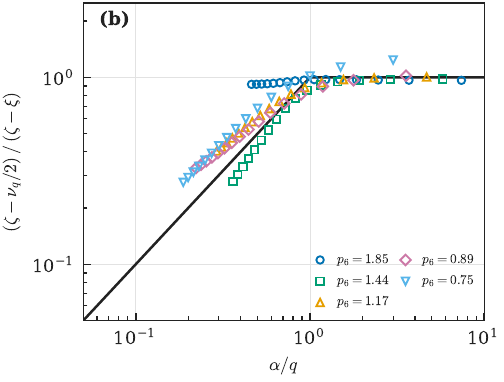}
  \caption{(\textbf{a}) The parameter dependence of the MSD exponent $\zg_2$ for the SDCA, and (\textbf{b}) a plot striving
  for a data collapse of all moments on a universal scaling curve. The parameter values $p_6$ for the right panel are
  specified in its legend. The solid lines show the analytical results of \eqns{sdcaMSD} and \eq{momentDataCollapse},
  respectively; the squares in the left panel were obtained from power-law fits of the MSD, at eleven values of $D$ that span
  both the strong anomalous phase and the Gaussian phase, $p_6 \geq 2$, where $\zg_2 = 1$. Details on the numerical
  parameters are provided in \secref{SDCA}.}
  \label{fig:identifyParameters}
\end{figure}

In applications one is often interested in the scaling of the MSD, $\moment = 2$. \figref{identifyParameters}a shows this
data for the SDCA of \secref{SDCA}, for which $\zeta = 3/2$ and $\moment = 2$ lies above the crossover moment throughout the
strong anomalous phase, so that $\zg_2 = 2\zeta - \zb$ takes the superballistic values of \eqn{sdcaMSD}, up to $\zg_2 \to 3$
as $p_6 \to 2/3$. It is then tempting to rewrite \eqn{momentPiecewise} in the form of a universal scaling curve that admits a
data collapse for all moments and all parameter values of any model showing strong anomalous diffusion with two linear
regimes of the spectrum:
\begin{align} \label{eq:momentDataCollapse}
  \frac{\zeta - \frac{\zg_\moment}{2}}{\zeta - \theta}
  = \left\{
  \begin{array}{lrl}
    \za / \moment & \quad\text{for} & \za / \moment < 1 \, ,
    \\[2mm]
    1 & \quad\text{for} & \za / \moment \geq 1 \, .
  \end{array}
  \right.
\end{align}
\figref{identifyParameters}b shows this representation for a collection of moments obtained by fitting SDCA data at five
values of $p_6$ (as discussed in \secref{SDCA}). The agreement is only approximate, in spite of clean numerical data: the
fitted exponents deviate systematically from the universal curve, most strongly for the large moments, which are dominated by
rare, extreme excursions, and for the moments closest to the crossover. These systematic errors are quantitatively explained
by the corrections to scaling derived in Ref.~\cite{arxiv} and recalled in \appref{fitProblems}: every moment is the
superposition of two power laws, \eqn{momentQ}, and the subleading one biases power-law fits up to the crossover time
\eq{critTimeScale}, which grows without bounds as $\moment \to \za$, and becomes very large whenever $\zeta - \theta$ is small.
Precisely where strong anomalous diffusion is most delicate, fitting individual moments is therefore unreliable, and the
exponents are better determined from the PDF as a whole: the scheme of Ref.~\cite{arxiv} that does so, recalled in
\appref{parameterFitting}, collapses the bulk and the tails of the distribution simultaneously, and it is put to work on the
SDCA in \secref{SDCA}.

\section{Specific Models}
\label{sec:discussion}

We test the theory on three model systems featuring strong anomalous transport: the SDCA in
shallow optical lattices \cite{MEZ96,KB12,AKB17}, the Slicer Map (SM) \cite{SRGK15}, and L\'evy walks (LW)
\cite{Metzler,ZDK15}. They feature different values of the exponents $\theta$ and $\za$, and they cover complementary
situations: an experimentally realized Langevin dynamics with unbounded velocities and a non-ballistic cutoff exponent,
$\zeta = 3/2$ (SDCA), an exactly solvable deterministic map (SM), and the classical annealed random walk with heavy-tailed
flights, both with $\zeta = 1$ (LW). For reference, \tabref{Spectrum} also summarizes the exponents reported in
Ref.~\cite{arxiv}: the L\'evy--Lorentz gas (LLg) \cite{BCV10,SRGK15}, the Lorentz gas with infinite (LGi) \cite{AHO03} and
finite (LGf) \cite{D14} horizon, and polygonal billiard channels (PBC) \cite{SL06,JR06,SS06,Orchard21}. In the LGf, the tails
of the distribution decay faster than any power law ($\za = \infty$), such that it cannot feature strong anomalous transport
(cf.~\appref{fastCutoff} and \secref{gaussian}).

\begin{table}[H]
  \caption{Choices of $\theta$, $\zeta$, and $\za$ that match the spectrum of exponents for the different model systems,
  characterizing the anomalous scaling of the distribution, \eqn{ProbScalingCutoff}, the algebraic tails of this
  distribution, \eqns{ProbTailZ} and \eq{ProbTailV}, and the time scaling of the cutoff function, \eqn{betadef}. The acronyms
  stand for: LLg, L\'evy--Lorentz gas; SM, Slicer Map; LW, L\'evy walks; LGi and LGf, Lorentz gas with infinite and finite
  horizon, respectively; PBC, polygonal billiard channels; SDCA, spatial diffusion of cold atoms. Irrespective of the choices
  adopted in the literature, we denote the parameters used to define the dynamics as $p_1, \dots, p_6$ (see
  \secref{discussion}). Adapted from Ref.~\cite{arxiv}.}
  \label{tab:Spectrum}
  \begin{tabular}{llccc}
    \toprule
    \textbf{Model} & \textbf{Parameter} & $\boldsymbol{\theta}$ & $\boldsymbol{\zeta}$ & $\boldsymbol{\za}$ \\
    \midrule
    LLg      & $0 < p_1 < 1$        & $(1+\za)^{-1}$ & $1$   & $p_1$      \\
             & $1 < p_1$            & $1/2$          & $1$   & $2p_1 - 1$ \\
    SM       & $0 < p_2$            & $0$            & $1$   & $p_2$      \\
    LW       & $1 < p_3 < 2$        & $p_3^{-1}$     & $1$   & $p_3$      \\
    LGi      & $0 < p_4 < 1$        & $1/2$          & $1$   & $2$        \\
    LGf      & $1 < p_4$            & $1/2$          & $1$   & $\infty$   \\
    PBC      & $p_5 \approx 1.0649$ & $\approx 0.5$  & $1$   & $\approx 2$ \\
             & $p_5 \approx 0.935$  & $\approx 0.5$  & $1$   & $\approx 3$ \\
    SDCA     & $1 < p_6 < 2$        & $p_6^{-1}$     & $3/2$ & $p_6$      \\
    \bottomrule
  \end{tabular}
\end{table}

\subsection{Spatial Diffusion of Cold Atoms}
\label{sec:SDCA}

The SDCA describes the motion of an atom in a shallow optical lattice under Sisyphus laser cooling. In the semiclassical
approximation \cite{MEZ96,AKB17}, the trajectory of an atom starting at the origin with $v(0) = 0$ is determined by the
Langevin equations
\begin{align}\label{eq:sdcaModel}
  \dot v(t) = - \frac{v}{1 + v^2} + \sqrt{2 D}\, \Gamma(t) \, ,
  \qquad
  \dot x(t) = v(t) \, ,
\end{align}
written in the dimensionless units of Ref.~\cite{AKB17}, where $\Gamma$ is Gaussian white noise and the momentum diffusion
constant $D$, proportional to the ratio of the recoil energy and the lattice depth, is the control parameter. The cooling
force is harmonic for $|v| \ll 1$ but decays as $-1/v$ for $|v| \gg 1$: fast atoms are only weakly damped, and perform long,
nearly free flights. For $D < 1$ the marginal velocity distribution reaches the power-law equilibrium
$P_{\rm eq}(v) \sim |v|^{-1/D}$ \cite{KB10,AKB17}, and in the range $1/5 < D < 1$ the displacement features strong anomalous
diffusion \cite{DL12,KB12,AKB17}, with
\begin{align}\label{eq:sdcaNu}
  \za = p_6 = \frac{1}{3D} + \frac{1}{3} \, ,
  \qquad
  \theta = \frac{1}{p_6} \, ,
  \qquad
  \zeta = \frac{3}{2} \, .
\end{align}
The bulk of the packet obeys the L\'evy scaling $r \sim t^{1/p_6}$ \cite{KB12}, while the tails are populated by atoms whose
velocities have themselves diffused up to $v \sim t^{1/2}$, giving displacements $r \sim t^{3/2}$: a Richardson-like scaling
that makes the SDCA the only system of \tabref{Spectrum} whose cutoff exponent is neither diffusive nor ballistic---so that
$\zeta$ must be fitted. The non-normalizable limit function $\lim_{t\to\infty} t^{\zeta+\zb}\, \mathcal P(r, t)$ is
the infinite covariant density of Refs.~\cite{KB10,AKB17}, whose exponent $1 + 1/(2D)$ coincides with our
$\zeta + \zb = 3 p_6/2 + 1/2$, by the hyper-scaling relation \eqn{betadef}. For the SDCA, \eqn{momentPiecewise} entails, in
the strong anomalous phase,
\begin{align}\label{eq:sdcaMSD}
  \zg_2 = 4 - \frac{3\, p_6}{2}
  \qquad \text{for } \; \frac{2}{3} < p_6 < 2 \, ,
\end{align}
and $\zg_2 = 1$ for $p_6 \geq 2$ ($D \leq 1/5$, the Gaussian phase \cite{AKB17}). Remarkably, $\zg_2$ exceeds the ballistic
value $2$ for $p_6 < 4/3$: the MSD is superballistic, although the bulk of the packet spreads subballistically,
with $\theta = 1/p_6 < 1$---an extreme form of bi-scaling (\figref{identifyParameters}a; cf.~\secref{checkMoments}).

We integrated \eqn{sdcaModel} with the Euler--Maruyama scheme and time step $0.01$, as in Ref.~\cite{AKB17}, for an ensemble
of $10^5$ trajectories up to time $10^5$ at $D = 0.4$, \ie{} $p_6 = 7/6$, the value of Fig.~1 of Ref.~\cite{AKB17}, and for
ensembles of $3 \times 10^4$ to $10^6$ trajectories, with time horizons between $10^3$ and $10^5$, at ten further values of
$D \in [0.15, 0.8]$, spanning the strong anomalous phase and reaching into the Gaussian phase. Horizons and ensemble sizes
are tied to each other by the requirement that the rare flights remain statistically resolved: since the fraction of ensemble
members populating the tail branch decays as $g(t) = t^{-\zb}$ (\eqn{betadef}), an ensemble of $N$ trajectories tracks the
flight branch of the moments only as long as $N t^{-\zb} \gg 1$, beyond which the sampled moments sag towards the bulk
scaling---a finite-sample counterpart of the crossover time \eq{critTimeScale}. All moment exponents are therefore fitted
over the latest window of one and a half decades satisfying $N t^{-\zb} \geq 30$, and the smallest values of $D$ in the
strong anomalous phase, which carry the largest $\zb = 3 p_6/2 - 1$, are assigned the largest ensembles over the shortest
horizons.

\begin{figure}[H]
  \includegraphics[width=0.48\linewidth]{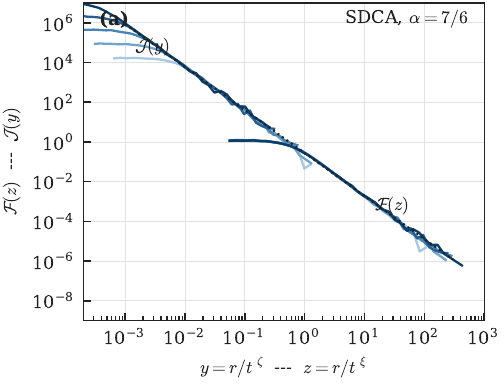}\hfill
  \includegraphics[width=0.48\linewidth]{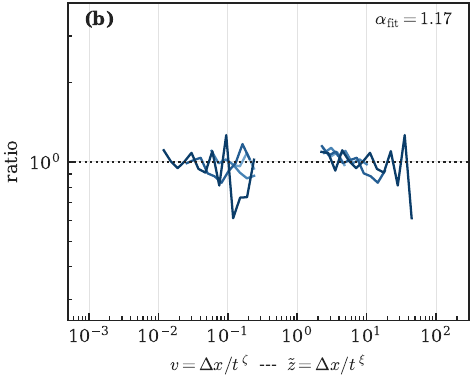}
  \\[2mm]
  \includegraphics[width=0.48\linewidth]{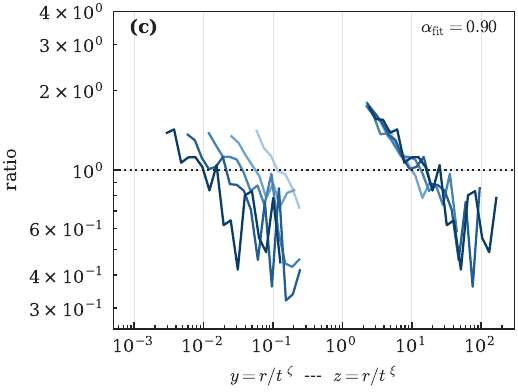}\hfill
  \includegraphics[width=0.48\linewidth]{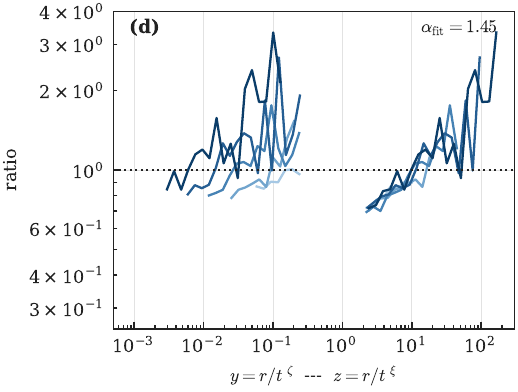}
  \caption{Simultaneous scaling of the bulk and of the tails of the PDF of the SDCA for the data also shown in
  \figref{PDFscaling}, \ie, for $D = 0.4$, where one expects $\za = p_6 = 7/6$, $\theta = 1/p_6 = 6/7$, $\zeta = 3/2$, and
  $\zb = \za\,(\zeta-\theta) = 3p_6/2 - 1 = 3/4$. (\textbf{a})~Joint plot of the data of both panels of \figref{PDFscaling}.
  (\textbf{b})~Reduced plot showing the ratio of the data and the pertinent values of the power law given by the dotted line
  in panel~(a). Panels (\textbf{c},\textbf{d}) show reduced plots of the best fits obtained when one rather adopts the values
  $\za = 0.90$ and $\za = 1.45$, respectively. Note that adopting a different value of $\za$ alters the power of $t$ in the
  definition of $\mathcal J(y)$, and the slope of the dotted line in panel~(a). Details on the model and on the numerical
  parameter values are provided in \secref{SDCA}.}
  \label{fig:collapseTest}
\end{figure}

The joint collapse of $\mathcal F$ and $\mathcal J$ for the main run is shown in \figrefs{PDFscaling} and
\fignum{collapseTest}: the two branches match on the intermediate power law with exponent $-(1+p_6) = -13/6$, and the reduced
plots single out $\za = p_6$ against the deliberately wrong values of \figref{collapseTest}c,d. The bulk-plateau fit of
\appref{parameterFitting} yields $\theta = 0.845 \pm 0.006$ against the expected $1/p_6 = 6/7 \simeq 0.857$; a linear fit of the
measured high-moment exponents, $\moment\,\zg_\moment/2 = \zeta \moment - \zb$ for $\moment \in [2.5, 5]$, returns
$\zeta = 1.43 \pm 0.04$ and $\zb = 0.76 \pm 0.05$ against the predicted $3/2$ and $3 p_6/2 - 1 = 3/4$, the uncertainties
reflecting the dependence on the admissible fit window. The hyper-scaling estimate $\za = \zb/(\zeta - \theta) = 1.3 \pm 0.2$ is
compatible with the crossover moment $p_6 = 7/6 \simeq 1.17$, with an accuracy limited by the slowly decaying corrections of
\eqn{momentScaling}: as anticipated in \appref{fitProblems}, the PDF-based reduced plots of \figref{collapseTest} constrain
$\za$ far more sharply than the moment fits do.

The scatter of the data in \figref{identifyParameters}b is due to the corrections to the power laws reported in
\eqn{momentScaling}. For the SDCA the relations $\zeta = 3/2$ and $\theta = 1/p_6$ imply
\begin{align} \label{eq:sdcaTau}
  \tau
  = \frac{1}{|\moment - \za| \, (\zeta - \theta)}
  = \frac{2\, p_6}{(3 p_6 - 2)\, |\moment - p_6|} \, ,
\end{align}
so that the correction terms decay slowly when $\moment \approx p_6$, and, at every $\moment$, when $p_6$ approaches $2/3$,
where the bulk and tail scalings merge. Consistently, in \figref{identifyParameters}b the largest deviations from the
universal curve occur for the moments closest to the crossover.

\subsection{The Slicer Map}
\label{sec:SM}

The Slicer Map was introduced in Ref.~\cite{SRGK15} as an analytically solvable dynamical system that features all regimes of
anomalous transport, from localization to ballistic motion, and it was shown to exhibit strong anomalous transport. Its
displacement autocorrelation function was obtained analytically \cite{GRTV17} and, more recently, a universal scaling of the
moments of displacement and of their correlation functions was established in Ref.~\cite{VRTGM21}. From a coarse-grained
perspective that focuses on the scaling of the distribution of the displacement, the SM is characterized as follows: the
initial distribution is an approximation of a delta function at the origin, and transport proceeds in a one-dimensional space
where the displacement from the origin is denoted by $r$. Each member $x$ of the ensemble is characterized by a life time
$t_c(x)$. For $t < t_c(x)$ it propagates with unit velocity; then it stops and remains at its terminal position $r = t_c(x)$,
up to negligible fluctuations, for all times $t > t_c(x)$. The distribution of life times obeys a power law,
$\operatorname{Prob}(t_c > t) = (t/t_0)^{-p_2}$, where $t_0$ is the smallest admissible life time.

For $t > t_0$ the dynamics provides a PDF of displacements that can be written in the form of \eqn{ProbScalingCutoff} with
\begin{align*}
  \mathcal{F}(z) &= \left\{
  \begin{array}{lll}
    0
    & \text{ for }
    & z = r < t_0 \, ,
    \\[1mm]
    \displaystyle
    \frac{p_2}{t_0^{-p_2}} \: z^{-p_2-1}
    & \text{ for }
    & t_0 < z = r < t \, ,
  \end{array}
  \right.
  \\[1mm]
  g(t)\,\mathcal{J}(y) &= \left\{
  \begin{array}{ll}
    \displaystyle
    t^{-p_2} \, \frac{p_2}{t_0^{-p_2}} \: y^{-p_2-1}
    & \text{ for } \quad \frac{t_0}{t} < y = \frac{r}{t} < 1 \, ,
    \\[3mm]
    \displaystyle
    t^{-p_2} \, t_0^{\,p_2} \: \delta(y - 1)
    & \text{ for } \quad y = \frac{r}{t} = 1 \, ,
    \\[2mm]
    0
    & \text{ otherwise, }
  \end{array}
  \right.
\end{align*}
such that $\theta = 0$, $\zeta = 1$, and $\za = \zb = p_2$, in agreement with the hyper-scaling relation \eq{betadef}. In
Ref.~\cite{GRTV17} the MSD was used to identify the relation between the parameters of the SM and of the L\'evy--Lorentz gas,
and it was pointed out that this entails a matching of all other moments. From the perspective of our present results, this
is an immediate consequence of \eqn{momentsScaling}. We return to the SM with more general, non-algebraic slicer sequences in
\secref{smSurvival}.

\subsection{L\'evy Walks}
\label{sec:LW}

L\'evy walks are arguably the most elaborately studied model system featuring strong anomalous transport
\cite{1989BlumenZumofenKlafter,RDHB14a,ZDK15,DBSB15}. We consider one-dimensional LWs where particles move with unit velocity
along a line: all particles start at the origin and move in a given direction for a time sampled from a L\'evy distribution
$f(t) \propto p_3\, t^{-p_3-1}$ with $t \in [1,\infty)$, reversing direction at the end of each step with probability $1/2$. For $1<p_3<2$,
the different scaling of the bulk   ($\theta=1/p_3$) and of the tails ($\zeta=1$, $\beta=p_3-1$ and $\alpha=p_3$) of the PDF was first observed in this context, where the distribution
$\mathcal J(y)$ of \eqn{ProbV} has been denoted a non-normalizable density \cite{RDHB14a,RDHB14b,ZDK15,VBB19,VBB20}. The moments of 
displacement, reviewed in Refs.~\cite{Metzler,ZDK15,VBB19,VBB20},  follow the behavior anticipated in \eqn{momentPiecewise} with
$\za = p_3$, $\zeta = 1$, and $\theta = 1/p_3$. 
The bulk $\mathcal F(r/t^{\theta})$ perfectly matches the tail $g(t)\, \mathcal J(r/t^{\zeta})$; in
particular, $\beta=\alpha(\zeta-\theta)$.

We simulated an ensemble of $2 \times 10^5$ trajectories up to time $10^5$ for $p_3 = \za = 1.5$ (data not shown). The joint
collapse of $\mathcal F$ and $\mathcal J$ robustly recovers the exponents of \tabref{Spectrum} together with the analytical
prefactors---in particular, the bulk-plateau fit yields $\theta = 0.665 \pm 0.003$, against the expected $1/p_3 = 2/3$--- while
reduced plots of the moments reveal that, even for such high-quality data, it is challenging to determine the asymptotic
scaling of the moments by fitting power laws: the corrections to scaling of \eqn{momentScaling} remain visible up to the
largest simulated times. The matching of the intermediate power laws of the bulk and of the tail, on the other hand, provides
robust results in the super-diffusive regime $1<p_3<2$.

\section{Gaussian Distributions}
\label{sec:gaussian}

The theory of \secref{moments} rests on the algebraic intermediate asymptotics \eq{ProbTailX}. It is natural to ask what
remains of the two-branch structure \eq{ProbScalingCutoff} when the distribution features no power-law regime---most
importantly, when the scaling function is Gaussian. In this section we give four complementary answers: a Gaussian scaling
function produces a single-linear spectrum of moments (\secref{GaussMono}); no non-algebraic shape can play the role of the
intermediate asymptotics, because asymptotic matching by itself forces a power law (\secref{GaussMono});
ballistic excursions that are exponentially rare, rather than power-law distributed, produce only a transient
imitation of strong anomalous diffusion, which can be recognized as such (\secref{transient}); and power-law
flights over a
Gaussian bulk yield \emph{bona fide} strong anomalous diffusion, with a crossover moment decoupled from the tail
exponent (\secref{crossRule}). Exactly solvable deterministic realizations of all these scenarios are constructed in
\secref{solvable}.

\subsection{Gaussian Scaling Functions Imply Single-Linear Spectrum}
\label{sec:GaussMono}

Consider the ansatz \eq{ProbScalingCutoff} with a Gaussian scaling function for the bulk, normalized on the half line,
\begin{align} \label{eq:GaussF}
  \mathcal{F}(z) = \sqrt{\frac{2}{\pi\sigma^2}} \; \exp\!\left( -\frac{z^2}{2\sigma^2} \right) ,
  \qquad z \geq 0 \, .
\end{align}
The Gaussian decays faster than any power law, so that the assumptions of \appref{fastCutoff} are satisfied, and all moments
are governed by the bulk branch:
\begin{align}\label{eq:GaussMoments}
  \left\langle r^\moment \right\rangle_t
  = N_\moment \, t^{\moment\,\theta}
  \qquad \text{with} \qquad
  N_\moment = \int_0^\infty z^\moment \, \mathcal{F}(z) \, \rmd z
  = \bigl( \sqrt{2}\,\sigma \bigr)^{\moment} \; \frac{\Gamma\!\left(\frac{\moment+1}{2}\right)}{\sqrt{\pi}} \, .
\end{align}
Hence $\moment\,\zg_\moment/2 = \theta\,\moment$ for \emph{all} moments: the spectrum is single-linear, transport is scale
invariant, and, in the language of \tabref{Spectrum}, $\za = \infty$, as for the Lorentz gas with finite horizon. For
$\theta = 1/2$ one recovers normal diffusion; for $\theta \neq 1/2$ one obtains weak anomalous diffusion with Gaussian statistics,
as realized, for instance, by fractional Brownian motion with Hurst exponent $\theta = H \in (0,1)$ \cite{MVN68}. In this case
no crossover moment exists, no logarithmic correction of the type of \eqn{momentAlpha} arises at any $\moment$, and the MSD
alone determines the complete spectrum of moments.

\subsection{Rare Ballistic Flights and Transient Bi-Scaling}
\label{sec:transient}

Consider a diffusive Gaussian background coexisting with ballistic flights that survive scattering for exponentially
distributed times. With unit speed ($\zeta = 1$) and $\theta = 1/2$, a minimal model is

\begin{align}\label{eq:GBmodel}
  \mathcal{P}(r, t)
  =  W(t) \; \delta( r - t ) +
  \bigl[ 1 - W(t) \bigr] \; t^{-\theta} \, \mathcal{F}\!\left( \frac{r}{t^{\theta}} \right)
  \, ,
  \qquad
  W(t) = \mathrm{e}^{-t/\tau} \, ,
\end{align}
Equation \eq{GBmodel} prescribes only the one-time distribution of the displacement: it selects no dynamics, and
infinitely many evolutions share these marginals.\footnote{A process is determined by its joint distributions at
multiple times, of which \eq{GBmodel} fixes only the single-time ones. Since the theory of \secref{moments}
involves nothing but $\mathcal P(r,t)$, its conclusions hold simultaneously for every such realization, stochastic
or deterministic; multi-time observables, such as the displacement autocorrelations studied in
Ref.~\cite{GRTV17}, do distinguish them. The fly-and-diffuse model of \secref{fad}, for instance, is fully
deterministic; it approaches these marginals only at late times, and \eq{GBmoments} describes its moments only in
that regime.} This structure---$\delta$ peaks at the ballistic front $r = t$, the largest displacement
reachable at the finite speed of the dynamics, with exponentially decaying weight, plus a smooth part converging
to a Gaussian---arises, for instance, in the telegraph (Goldstein--Kac) process, in which a particle moves at
constant speed and reverses its direction at exponentially distributed times \cite{Goldstein51,Kac74}. Truncating the Gaussian at $r \leq t$ does not affect the asymptotic
statements below. The moments read:
\begin{align}\label{eq:GBmoments}
  \left\langle r^\moment \right\rangle_t
  = \mathrm{e}^{-t/\tau} \; t^{\moment\,\zeta} +
  \bigl[ 1 - W(t) \bigr] \, N_\moment \, t^{\moment\,\theta}
  \, ,
\end{align}
with $N_\moment$ as in \eqn{GaussMoments}: the same two-power-law structure as \eqn{momentQ}, except that the amplitude of
the flight term decays \emph{faster than any power} $t^{-\zb}$, consistently with Proposition~\ref{prop:rigidity}. This
difference has sharp consequences.

\begin{figure}[H]
  \includegraphics[width=\linewidth]{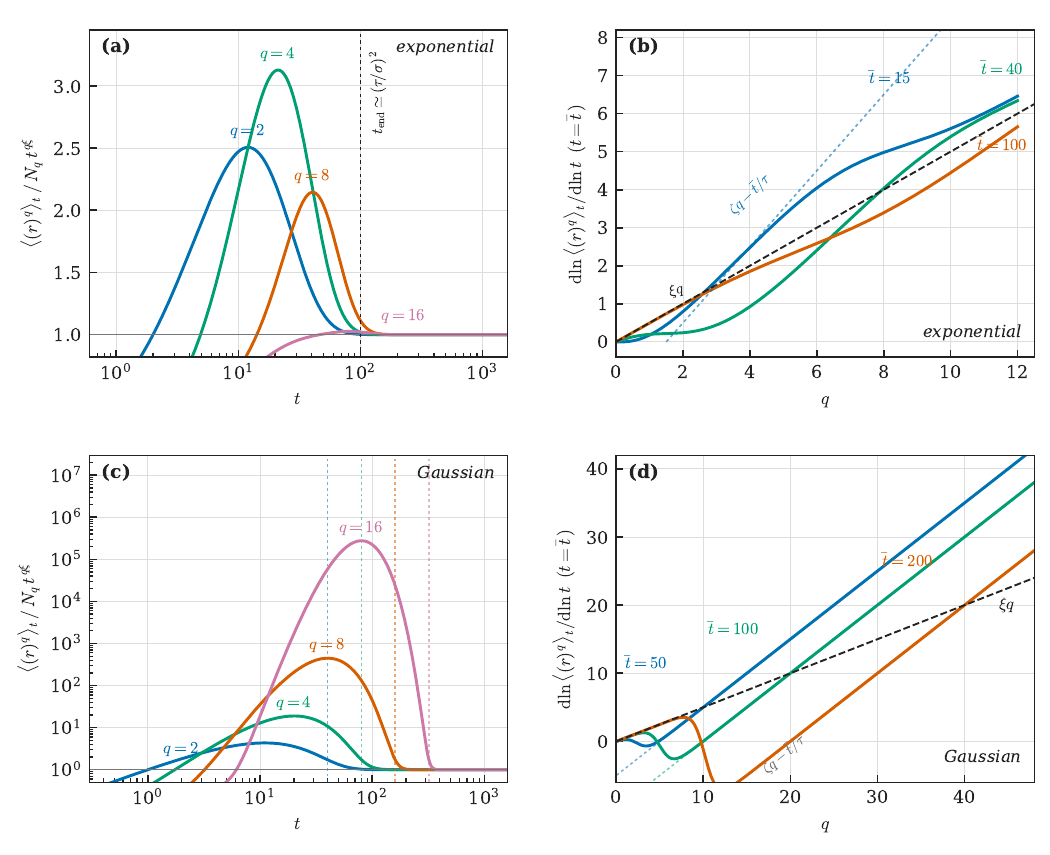}
  \caption{Transient imitation of strong anomalous diffusion in the minimal model \eq{GBmodel}, evaluated from the exact
  expression \eq{GBmoments} with $\sigma = 1$, $\tau = 10$, $\theta = 1/2$, $\zeta = 1$ (analytics, no simulation), for an
  exponential (top row) and a Gaussian (bottom row) scaling function. (\textbf{a})~Exponential case, reduced moments
  $\langle r^\moment \rangle_t / N_\moment t^{\moment\theta}$ for $\moment = 2, 4, 8, 16$: the humps are bounded, and they all
  terminate by the closure time $t_{\rm end} \simeq (\tau/\sigma)^2$ of \eq{tEnd} (dotted vertical line); moments with
  $\moment \gtrsim \tau/\sigma^2$ are never dominated by the flight term. (\textbf{b})~Exponential case, apparent spectrum at
  $\bar t = 15$, $40$, $100$: the flight branch (dotted line: $\zeta\moment - \bar t/\tau$ for $\bar t = 15$) is a bounded
  bubble that rejoins the bulk line $\theta\moment$ (dashed) at large $\moment$, and it disappears for $\bar t > t_{\rm end}$.
  (\textbf{c})~Gaussian case, reduced moments: each moment displays a hump, peaking at
  $t \approx \moment\,(\zeta-\theta)\,\tau$, before settling on the plateau at the crossover time $t_\moment \simeq a\,\moment$
  (dotted vertical lines), \eqn{aDef}. (\textbf{d})~Gaussian case, apparent spectrum at $\bar t = 50$, $100$, $200$: the
  flight branches $\zeta\moment - \bar t/\tau$ (dotted) drift linearly with $\bar t$, and so does their knee with the bulk
  line $\theta\moment$ (dashed), unlike the time-independent spectrum \eq{momentPiecewise} of genuine strong anomalous
  diffusion. The negative excursion of the $\bar t = 200$ curve is an exact feature of the minimal model, discussed in the
  text.}
  \label{fig:GaussTransient}
\end{figure}

Firstly, for every fixed $\moment$ the flight term is asymptotically subdominant: as $t \to \infty$ the spectrum is
single-linear, in agreement with \secref{GaussMono}. Equating the logarithms of the two terms in \eqn{GBmoments}, and using
$\ln N_\moment \simeq (\moment/2)\,[\ln(\sigma^2 \moment) - 1]$ for large $\moment$, one finds that the crossover structure
is governed by a single constant $a$: the flight term dominates the value of the $\qmom$ moment up to the time
\begin{subequations}\label{eq:aDef}
\begin{align}
  t_\moment \simeq a\, \moment \, ,
  \qquad\text{and, at fixed $t$, for moments } \moment \gtrsim \tilde{\moment}(t) \simeq \frac{t}{a} \, ,
\end{align}
where $a$ solves
\begin{align}
  a = \frac{\tau}{2} \left[ 1 + \ln \frac{a}{\sigma^2} \right]
  \qquad (\theta = 1/2, \; \zeta = 1) \, .
\end{align}
\end{subequations}
Such a dominance window exists only if $\tau > 2\sigma^2$ (in units of unit speed), as one verifies by maximizing over
$\moment$ at fixed $t$: rare flights leave a visible trace in the moments only when their persistence time exceeds the
diffusive scale of the background. For the telegraph process itself, $\sigma^2 = \tau$ at unit speed, so the criterion fails,
and the flight term never dominates at late times; pronounced transients require flights that live much longer than the bulk
scattering time. For $\tau = 10\,\sigma^2$ one finds $a \simeq 20\,\sigma^2$, and the numerically determined dominance times
$t_\moment = \{36, 76, 155, 315\}$ for $\moment = \{2,4,8,16\}$ closely follow $t_\moment \simeq a \moment$; see
\figref{GaussTransient}c.

Secondly, and most importantly for data analysis, consider fitting power laws over an observation window centered at a time
$\bar t$. The local logarithmic slope of the flight term is
\begin{align} \label{eq:apparentBranch}
  \frac{\rmd \ln \bigl[ \mathrm{e}^{-t/\tau}\, t^{\moment\zeta} \bigr]}{\rmd \ln t}
  = \zeta \moment - \frac{t}{\tau} \, ,
\end{align}
so that the moments dominated by it display an apparently \emph{linear} second branch, $\zeta\moment - \zb_{\text{eff}}$,
with effective offset $\zb_{\text{eff}} = \bar t/\tau$. An observer interpreting such data through \eqn{momentPiecewise}
would extract an apparent crossover moment $\za_{\text{eff}} = \zb_{\text{eff}}/(\zeta - \theta) = \bar t / [\tau (\zeta-\theta)]$,
and the fitted pair $(\za_{\text{eff}}, \zb_{\text{eff}})$ would even satisfy the hyper-scaling relation \eq{betadef}. At any
single observation time, therefore, the transient scenario is indistinguishable from genuine strong anomalous
diffusion---even by the hyper-scaling check. The give-away is the \emph{drift}: both $\za_{\text{eff}}$ and
$\zb_{\text{eff}}$ grow linearly with the observation time, whereas for power-law tails they are constants.
\figref{GaussTransient}d displays the apparent spectrum of the minimal model at $\bar t = 50$, $100$, and $200$: the knee
between the bulk line $\theta\moment$ and the drifting branches $\zeta\moment - \bar t/\tau$ moves to the right proportionally
to $\bar t$. Between $\tilde\moment(\bar t)$ and the knee, the local exponents even sag below the bulk line: these are the
descending flanks of the humps of \figref{GaussTransient}c, where the flight term still dominates the value of the moment
while already decaying. For $\bar t/\tau$ large enough, the sag drives the local exponents below zero---the $\bar t = 200$
curve of \figref{GaussTransient}d is negative precisely for $\tilde\moment(\bar t) \lesssim \moment < \bar t/\tau$: in the
minimal model the mass leaving the flight term is redistributed instantaneously over the whole bulk, so that these moments
transiently \emph{decrease}. This is an artifact of that idealization. In the particle realizations of \secref{smSurvival}
and \secref{fad} the moments never decrease: for the slicer class,
$\rmd \langle r^\moment \rangle_t / \rmd t = \moment\, t^{\moment-1}\, W(t) > 0$ exactly, and in the fly-and-diffuse model
the stopped mass populates the region just behind the front before spreading, so that the measured exponents sag toward, but
not below, zero.

The Gaussian bulk is, in fact, the marginal member of a whole family. For an \emph{exponential} scaling function,
$\mathcal F(z) = \sigma^{-1}\, \mathrm{e}^{-z/\sigma}$, the bulk amplitudes $N_\moment = \sigma^\moment\, \Gamma(\moment+1)$
grow as $\mathrm{e}^{\moment \ln \moment}$, twice as fast, in the exponent, as in the Gaussian case \eq{GaussMoments}, and
the flight term loses its niche within a finite time: maximizing the flight-to-bulk ratio over $\moment$ shows that the
dominance window closes, up to subleading corrections, at
\begin{align} \label{eq:tEnd}
  t_{\rm end} \simeq \left( \tau/\sigma \right)^{2} ,
\end{align}
and that moments with $\moment \gtrsim \tau/\sigma^2$ are never dominated by the flights at all. The humps of the reduced
moments are then bounded, and they all terminate by $t_{\rm end}$ (\figref{GaussTransient}a). The apparent second
branch---which retains the local slope $\zeta\moment - \bar t/\tau$ of \eq{apparentBranch}, since the latter is a property of
the flight term alone---becomes a bounded bubble that rejoins the bulk line at large $\moment$, and it disappears altogether
for $\bar t > t_{\rm end}$ (\figref{GaussTransient}b): with an exponential bulk, the imitation of strong anomalous diffusion
does not drift, it dies.

\begin{Remark}\label{rem:stretched}
For stretched-exponential scaling functions, $\mathcal F(z) \propto \exp\left( -z^\gamma / \gamma \sigma^\gamma \right)$, one
has, for large $\moment$, $\ln N_\moment \simeq (\moment/\gamma) \left[ \ln(\sigma^\gamma \moment) - 1 \right]$, and
the same maximization yields, for $\theta = 1/2$ and $\zeta = 1$, a dominance window that closes at
\begin{align*}
  t_{\rm end} \simeq \left[ \frac{\tau}{\gamma\, \sigma^{\gamma}} \right]^{2/(2-\gamma)}
  \qquad \text{for } \gamma < 2 \, ,
\end{align*}
while for $\gamma > 2$ the window is present at all sufficiently late times. The Gaussian case, $\gamma = 2$, is exactly
marginal: the exponent $2/(2-\gamma)$ diverges, and the existence of the window is decided by the time-independent criterion
$\tau > 2\sigma^2$ established above. The thinner the tail of the bulk, the longer-lived the imitation.
\end{Remark}

In practice, three tests distinguish genuine strong anomalous diffusion from its finite-time imitation. (i)~\emph{Stability:}
upon doubling the observation time, the fitted crossover moment of genuine strong anomalous diffusion is stable, while in the
transient scenario it doubles (Gaussian bulk, cf.~\eqn{apparentBranch}) or the second branch disappears altogether (heavier
bulk tails, Remark~\ref{rem:stretched}). (ii)~\emph{Collapse:} the data analysis of \appref{parameterFitting} collapses the
tail branch with a single, time-independent exponent $\zb$ only for power-law tails; for an exponentially damped
flight term, $t^{\zeta+\zb}\,\mathcal P$ at fixed $y$ is proportional to $t^{\zb}\, \mathrm{e}^{-t/\tau}$, so the curves at
successive times sag for \emph{every} choice of $\zb$, and no collapse is possible. (iii)~\emph{Crossover times:} reduced
moments that reach their asymptotic plateaus at times growing linearly with the moment order, \eqn{aDef}, are the fingerprint
of an exponentially damped ballistic component, whereas in genuine strong anomalous diffusion the crossover time
\eq{critTimeScale} diverges only for $\moment$ near $\za$. In the Lorentz gas with finite horizon the flight lengths are
strictly bounded, so that there even the transient branch is cut off beyond a fixed time \cite{D14}.

This analysis reinforces the practical message of \appref{fitProblems} and \appref{parameterFitting}: an apparent bi-linear
spectrum at a single observation time does not, by itself, establish strong anomalous diffusion; the scaling analysis of the
PDF, and the stability of the fitted exponents under an increase of the observation time, do.

\subsection{A Gaussian Bulk with Power-Law Flights: the Crossover Rule}
\label{sec:crossRule}
The third possibility is a Gaussian bulk coexisting with flights whose survival is itself a power law: take
$W(t) = (1 + t/t_0)^{-\zb}$, with $\zb > 1$, so that the mean flight time is finite and the bulk spreads
diffusively, $\theta = 1/2$, $\zeta = 1$, and let each member, when its flight ends, diffuse from the position it
has reached. 
The marginals then carry \emph{three} components, not the two of \eq{GBmodel}: between the Gaussian bulk and the
ballistic front one finds the \emph{record} of the terminated flights. By this we mean that, at unit speed, a flight 
ending at time $s$ also ends at position $r = s$, so the death times are {\em recorded} as positions. The three regions
are then characterized by a probability density of the following form,  
where the middle component is new as compared to \eq{GBmodel}:
\begin{align*}
  \mathcal P(r, t) \simeq
  \left\{
  \begin{array}{lll}
    K(t) \, t^{-1/2}\, \mathcal F\!\left( r/t^{1/2} \right)
    & \text{ for } & 0 \le r < A(t)\, ,
    \\[1mm]
    -W'(r) \simeq \zb\, t_0^{\zb}\, r^{-(1+\zb)}
    & \text{ for } & A(t) < r < t \, ,
    \\[1mm]
    W(t)\, \delta(r - t)
    & \text{ at } & r = t \, ,
  \end{array}
  \right.
\end{align*}
where $A(t)$ is an intermediate scale, $t^{1/2} \ll A(t) \ll t$, and $K(t)$ is a suitable normalization
factor: since the mass beyond the bulk is exactly $\int_{A(t)}^\infty \mathcal P(r, t)\, \rmd r = W(A(t))$,
which vanishes as $t \to \infty$, one has $K(t) \to 1$.
By recasting $\mathcal P(r, t)$ in the form of the ansatz
\eq{ProbScalingCutoff}, as for the Slicer Map of \secref{SM}, one can write:
\begin{align*}
  \mathcal F(z) = \sqrt{\frac{2}{\pi\sigma^2}}\; \mathrm e^{-z^2/2\sigma^2} ,
  \quad
  g(t)\, \mathcal J(y) = t^{-\zb} \Bigl[ \zb\, t_0^{\zb}\, y^{-(1+\zb)} + t_0^{\zb}\, \delta(y-1) \Bigr]
  \quad \text{for } \; t^{\theta - \zeta} \ll y \le 1 \, ,
\end{align*}
with $z = r/t^{1/2}$ and $y = r/t$. Integrating $r^\moment$ against the three pieces, the bulk contributes
$N_\moment\, t^{\moment/2}$, the record $\zb\, t_0^{\zb}\, t^{\moment-\zb}/(\moment-\zb)$, and the front
$t_0^{\zb}\, t^{\moment-\zb}$, so that, to leading order,
\begin{align}\label{eq:crossMoments}
  \left\langle r^\moment \right\rangle_t
  \simeq N_\moment\, t^{\moment/2} + \frac{\moment}{\moment - \zb}\; t^{\moment}\, W(t)
  = N_\moment\, t^{\moment/2} + \frac{\moment\, t_0^{\zb}}{\moment - \zb}\; t^{\moment - \zb}
  \qquad \text{for } \moment > \zb \, , \; t \gg t_0 \, ,
\end{align}
where the record and the front combine into the first form, which holds for any survival function with regularly
varying tail of index $-\zb$---by the rescaling of \appref{slicerW}---and the second form evaluates it to leading
order in $t_0/t$. Below $\zb$, the record only adds a constant. Had the dead members instead been reinjected into
the bulk, as in \eq{GBmodel}, the record would be absent and the flight term would carry unit amplitude,
$t^\moment\, W(t)$: same exponents, same knee, but a different amplitude---the moments remember the bookkeeping.
Each moment is dominated by the faster-growing branch, and the knee sits where the two growth rates cross:
\begin{align}\label{eq:crossRule}
  \frac{\moment\, \zg_\moment}{2}
  = \max\!\left( \frac{\moment}{2} \, , \; \moment - \zb \right) ,
  \qquad
  \moment_c = \frac{\zb}{\zeta - \theta} = 2\, \zb \, .
\end{align}
In the matched scenario of \secref{moments} the overlap of the two branches forces, in addition, $\za = \moment_c$:
the crossover moment coincides with the exponent of the shared power law, and the hyper-scaling relation \eq{betadef}
expresses this coincidence. Here no such overlap exists---the Gaussian bulk has no power-law tail to share, realizing
the second alternative of Corollary~\ref{cor:noGauss}, in full consistency with Proposition~\ref{prop:rigidity}; and
\appref{fastCutoff} is not contradicted, because its hypothesis---faster-than-power decay beyond the bulk
scale---fails. The tail exponent is therefore fixed by the flight kinematics alone, and the crossover decouples from
it: $\moment_c = 2\zb$, twice the tail exponent for the diffusive bulk. Two signatures separate the scenarios. At
$\moment = \moment_c$ the matched case carries the logarithmic enhancement of \eqn{momentAlpha}, while here the two
amplitudes of \eq{crossMoments} simply add, and no logarithm appears. And since the pair $(\moment_c, \zb)$ satisfies
$\zb = \moment_c (\zeta - \theta)$ by construction, the hyper-scaling check alone cannot tell the two scenarios apart;
the scaling analysis of the PDF can. 
Finally, we notice that $\nu_2 = 1$: the MSD grows linearly in time, and yet the diffusion is strongly
anomalous.
An exact realization, amplitudes included, is constructed in \secref{gaussSAD}.

\section{Exactly Solvable Deterministic Models}
\label{sec:solvable}

The scenarios of \secref{gaussian} are realized exactly, amplitudes included, within the Slicer Map class of
\secref{SM} and its fly-and-die and fly-and-diffuse variants, and the modified maps can be written explicitly. The
coarse-grained dynamics of the whole family is specified by a single function, the survival probability $W(t)$ of
the ballistic flights, and each behavior corresponds to a choice of $W$: power laws yield strong anomalous
diffusion, faster-decaying survival functions its transient imitation, and a diffusing variant with power-law $W$ a
Gaussian bulk coexisting with power-law flights.

\subsection{Slicer Maps with Generic Survival Functions}
\label{sec:smSurvival}

The SM acts on the chain of unit cells $\hat{M} = [0,1] \times \mathbb{Z}$, with the ensemble initially
uniform in cell $m = 0$, as \cite{SRGK15}
\begin{align} \label{eq:SMdef}
  S(x, m) =
  \left\{
  \begin{array}{ll}
    (x,\, m-1) & \text{ for } \; 0 \le x < \ell_{|m|} \;\text{ or }\; \tfrac{1}{2} < x \le 1 - \ell_{|m|} \, ,
    \\[1mm]
    (x,\, m+1) & \text{ for } \; \ell_{|m|} \le x \le \tfrac{1}{2} \;\text{ or }\; 1 - \ell_{|m|} < x \le 1 \, ,
  \end{array}
  \right.
\end{align}
where the ``slicers'' $\ell_{|j|}$ and $1 - \ell_{|j|}$ cut cell $j$, and the original choice
$\ell_{|j|} = (|j| + 2^{1/p_2})^{-p_2}$ yields the power-law survival of \secref{SM}. Points travel ballistically until their
coordinate first falls on the inner side of a slicer, after which they bounce between two neighboring cells---the
coarse-grained ``death''. The mechanism only requires $\ell_0 = 1/2$, as in the original sequence, and monotonic decay.
The coarse-grained dynamics is thus specified entirely by the survival function $W(t)$---the fraction of the ensemble
still flying at time $t$, which is the survival probability $\operatorname{Prob}(t_c > t)$ of the life time $t_c$. For
the Slicer Map, we have
\begin{align*}
  W(j) = 2\, \ell_{|j|} \, ,
\end{align*}
whatever law describes $\ell_{|j|}$; $W$ may then be read as a step function of the continuous time $t$.
Equivalently, in the fly-and-die
formulation \cite{VRTGM21}---where the member of initial coordinate $x_0$, uniform on $(0,1]$, has position
$x(x_0, t) = x_0 + t$ for $t \le t_c(x_0)$, frozen at $x_0 + t_c(x_0)$ afterwards---the original life-time function is
$t_c(x_0) = t_0\, x_0^{-1/p_2}$. The probability that the life ends in $[t, t + \rmd t)$ is
$-\rmd W(t) = -W'(t)\,\rmd t$ when $W$ is differentiable; if it is not, like the step
function of the Slicer Map, $-\rmd W(t)$ is defined just as easily, as illustrated in \appref{slicerW}.
In a few words: \eq{SMdef} and the slicer sequences completely specify the map. Everything from \eq{SMgenC} to
\eq{SMgenBranches} involves only $W$, and holds verbatim for the continuous-time fly-and-die process, of
which the map is the integer-time sampling. This
characterization of the ensemble is summarized by the cumulative distribution of $R(t)$, the displacement at time $t$.
For a displacement value $r < t$, $R(t)$ does
not exceed $r$ precisely when the life ended by time $r$, so that
\begin{align} \label{eq:SMgenC}
  \operatorname{Prob}\left( R(t) \le r \right)
  = \left\{
  \begin{array}{lrl}
    1 - W(r) & \quad\text{for} & 0 \le r < t \, ,
    \\[1mm]
    1 & \quad\text{for} & r \ge t \, ,
  \end{array}
  \right.
\end{align}
which is the fraction $1 - W$ behind the front and jumps by the still-flying fraction $W(t)$ at $r = t$. For the sake of
illustration take $W$ differentiable; then differentiation with respect to $r$ yields the displacement PDF:
\begin{align} \label{eq:SMgenP}
  \mathcal{P}(r, t) = -W'(r)\, \Theta(t - r) + W(t)\, \delta(r - t) \, ,
\end{align}
where $\Theta$ is the Heaviside step function and the $\delta$ term---the derivative of the jump---carries the
still-flying fraction, cf.\ \appref{slicerW}.
Of course, $\delta$ is not a function: \eqn{SMgenP} is a formal but useful rewriting of
\eqn{SMgenC}, and multiplication by the volume element $\rmd r$ restores a bona fide probability element---the
life-time element, which is $-\rmd W(r)$ on $(0, t)$, plus an atom of mass $W(t)$ at the front---while every use made of
$\mathcal{P}$ below amounts to an integral against observables.

Observe that, at unit speed, position and death time coincide for every dead member: an integral over death
times is simultaneously an integral over the frozen positions, which fill exactly $(0, t)$, with unit Jacobian---the
special gift of the slicer class.
A detailed derivation of
\eqns{SMgenP}, \eq{SMgenM}, \eq{SMgenMln} and \eq{SMgenBranches} is provided in \appref{slicerW}. Since only the flying
members advance, all at unit speed, the moments obey
\begin{align*}
  \frac{\rmd \langle r^\moment \rangle_t}{\rmd t} = \moment\, t^{\moment-1}\, W(t)
\end{align*}
exactly, whence
\begin{equation}
\label{eq:SMgenM}
  \left\langle r^\moment \right\rangle_t
  = \int_0^t s^\moment \left( -W'(s) \right) \rmd s + t^\moment\, W(t)
  = \moment \int_0^t s^{\moment-1}\, W(s)\, \rmd s \, ,
\end{equation}
and
\begin{equation}
\label{eq:SMgenMln}
  \frac{\rmd \ln \left\langle r^\moment \right\rangle_t}{\rmd \ln t}
  = t \frac{\rmd \ln \left\langle r^\moment \right\rangle_t}{\rmd t}
  = \frac{\moment\, t^{\moment}\, W(t)}{\left\langle r^\moment \right\rangle_t} \, ,
\end{equation}
Introducing the running exponent of the survival function,
\begin{align*}
  \varepsilon(t) := -\frac{\rmd \ln W}{\rmd \ln t} \, ,
\end{align*}
the exact spectrum \eq{SMgenMln} takes the asymptotic form
\begin{align} \label{eq:SMgenBranches}
  \frac{\rmd \ln \left\langle r^\moment \right\rangle_t}{\rmd \ln t}
  \simeq \left\{
  \begin{array}{l}
    0
    \\[1mm]
    \moment - \varepsilon(t)
  \end{array}
  \right.
  \;\; = \;\;
  \left\{
  \begin{array}{lrl}
    \theta\, \moment & \quad\text{for} & \moment \lesssim \varepsilon(t) \, ,
    \\[1mm]
    \zeta\, \moment - \varepsilon(t) & \quad\text{for} & \moment \gtrsim \varepsilon(t) \, ,
  \end{array}
  \right.
\end{align}
with the exponents identified as in \secref{transient}. Here, $\zeta = 1$, because the surviving members fly ballistically,
so that the front $r = t$ is the moving cutoff, and $\theta = 0$---in place of the diffusive $1/2$---because the bulk is
now the frozen record $-W'(r)$, which does not spread at all: below the knee the moments saturate to constants, and
the lower branch of the spectrum is flat. The knee at
$\moment \approx \varepsilon(t)$ is smeared over a range of moments of order $\sqrt{\varepsilon}$---the width over which
the endpoint $s = t$ of the integral in \eqn{SMgenM} competes with the interior maximum of its integrand---and it is
therefore relatively sharp, since $\sqrt{\varepsilon}/\varepsilon \to 0$ as $\varepsilon$ grows.

\subsection{Stretched-Exponential Slicers: the Drifting Knee}
\label{sec:smStretched}

For the present purpose, the original law for the slicers can be simply modified with the single replacement
\begin{align} \label{eq:SMmod}
  \ell_{|j|} = \frac{1}{2} \; \mathrm{e}^{-(|j|/\tau)^{\gamma}} \, ,
\end{align}
which leaves \eq{SMdef}, the phase space, and the non-chaotic, measure-preserving character of the map untouched, and gives
the flying fraction $W(m) = 2\,\ell_m = \mathrm{e}^{-(m/\tau)^{\gamma}}$ after $m$ steps.
Analogously, the fly-and-die map requires
\begin{align} \label{eq:FnDmod}
  t_c(x_0) = \tau \left( \ln \frac{1}{x_0} \right)^{\!1/\gamma} ,
\end{align}
for which $\operatorname{Prob}(t_c > t) = \mathrm{e}^{-(t/\tau)^{\gamma}}$ holds exactly at all $t \ge 0$.

\begin{figure}[H]
  \includegraphics[width=\linewidth]{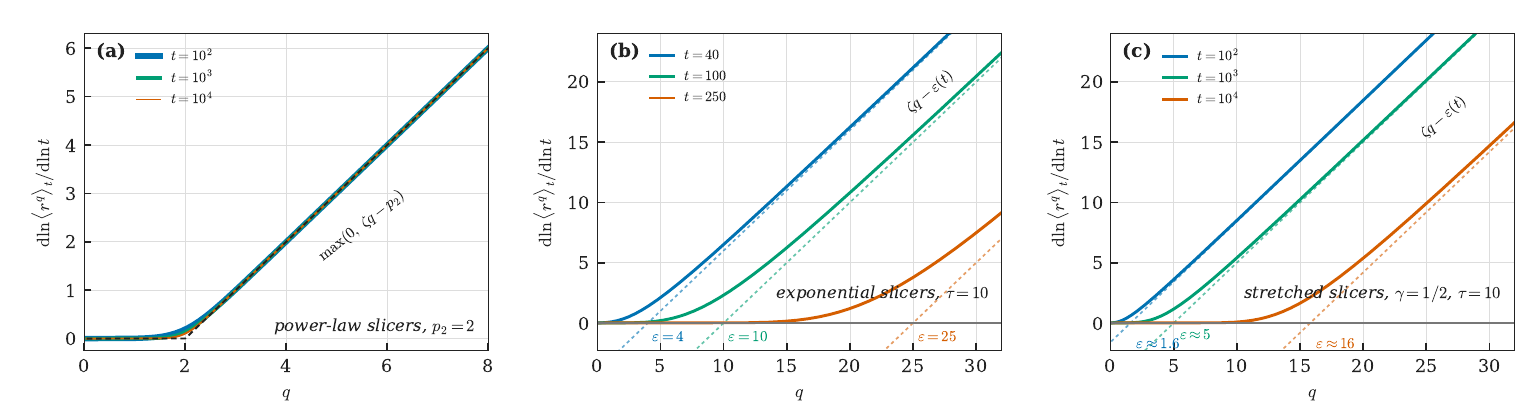}
  \caption{Exact spectra of the Slicer Map with generic survival functions, \eq{SMgenM}. (\textbf{a})~Power-law slicer
  sequence, $W = (t/t_0)^{-p_2}$ with $p_2 = 2$: the spectra at $t = 10^2$, $10^3$, $10^4$ superpose (drawn with decreasing
  line widths) on the
time-independent bi-linear form $\max(0, \zeta\moment - p_2)$ (dashed).
  (\textbf{b})~Exponential slicer sequence, $W = \mathrm{e}^{-t/\tau}$ with $\tau = 10$: the spectra at $t = 40$, $100$,
  $250$ follow the drifting branches $\zeta\moment - \varepsilon(t)$ (dotted), with $\varepsilon(t) = t/\tau = 4$, $10$, and
  $25$: the knee walks to the right as the observation time grows. (\textbf{c})~Stretched slicer sequence,
  $W = \mathrm{e}^{-(t/\tau)^{1/2}}$ with $\tau = 10$: at $t = 10^2$, $10^3$, $10^4$ the knee
  $\varepsilon(t) = \frac{1}{2} (t/\tau)^{1/2} \approx 1.6$, $5$, $16$ drifts sublinearly---two decades of time move
  it by about one of the exponential case's steps.}
  \label{fig:SlicerSurvival}
\end{figure}

\eqn{SMgenBranches} condenses the dichotomy of the present analysis into a single
statement: \emph{the apparent offset of the second
branch is the running exponent of the survival function}.
It is time independent---and the spectrum bi-linear at all times---if
and only if $W$ is a power law: for $W = (t/t_0)^{-p_2}$ one has $\varepsilon \equiv p_2$, and \eq{SMgenBranches} reproduces
the SM spectrum of \secref{SM}, with $\za = \zb = p_2$ (\figref{SlicerSurvival}a). For the stretched-exponential
sequence \eq{SMmod}, instead, $\varepsilon(t) = \gamma\, (t/\tau)^{\gamma}$: the knee drifts linearly in time for
exponential slicers ($\gamma = 1$,
\figref{SlicerSurvival}b), and quadratically for Gaussian ones ($\gamma = 2$), realizing at $\theta = 0$, and in closed form,
the two columns of \figref{GaussTransient}.
For $0 < \gamma < 1$, the properly stretched case, the knee drifts sublinearly, as $t^{\gamma}$ ($\gamma = 1/2$,
\figref{SlicerSurvival}c): the smaller $\gamma$,
the longer-lived the imitation.
All members of the family \eq{SMmod} with fixed $\gamma > 0$ are thus equivalent in kind, however different in
appearance: exponential, Gaussian and stretched slicers differ in the law of the knee's drift, but share the same
ultimate fate, the saturation of every moment---for each of them, the imitation is transient.

The equivalence breaks
only at the boundary of the family. In fact, \eqn{SMmod} interpolates continuously between these two scenarios:
as long as $\gamma \ln(t/\tau) \ll 1$, one has $\mathrm{e}^{-(t/\tau)^{\gamma}} \simeq \mathrm{e}^{-1}\, (t/\tau)^{-\gamma}$,
a power law with exponent $\gamma$, so that $\varepsilon(t) \simeq \gamma$ remains constant over the window
$t \ll \tau\, \mathrm{e}^{1/\gamma}$, which exceeds any accessible time scale for small $\gamma$. In that limit the
stretched-exponential slicer reduces to the power-law slicer with $p_2 = \gamma$: the knee does not drift, and genuine
strong anomalous diffusion is recovered.
At every fixed $t$ the apparent pair
$\za_{\text{eff}} = \zb_{\text{eff}} = \varepsilon(t)$ again satisfies the hyper-scaling relation \eq{betadef}, since
$\zeta - \theta = 1$. Furthermore,
since the frozen record does not grow at all while the front contributes $\sim t^{\,\moment - \varepsilon(t)}$,
the two branches compete at the knee at zero growth rate---the branch-value crossing of \secref{gaussSAD} is here
exactly marginal, the moment at the knee advancing only logarithmically---so that the crossing is decided by
logarithms rather than by powers of $t$. The knee is nevertheless sharp, its width, of order $\sqrt{\varepsilon}$,
remaining small compared with its drifting position $\varepsilon(t)$.

\subsection{The Fly-and-Diffuse Model: the Escaping Knee}
\label{sec:fad}

The slicer class realizes the scenarios of \secref{gaussian} with a frozen bulk, $\theta = 0$: asymptotically, transport stops
altogether. To realize the Gaussian minimal model \eq{GBmodel} itself---exponentially rare flights over a bulk that keeps
transporting---it suffices to let the members survive the end of their flight. In the \emph{fly-and-diffuse} model, a member
flies ballistically, at unit speed, for a life time drawn from the survival function $W(t) = \mathrm{e}^{-t/\tau}$, and then
continues with diffusive motion of diffusivity $D$. The diffusive stage can itself be generated by a
deterministic chaotic map \cite{GN82,GF82,Klages07}, keeping the dynamics fully deterministic. For the sake of
illustration, here we take the Baker Map, which is two-dimensional---the minimum necessary to be time-reversal
invariant---and preserves the Lebesgue measure. Like its numerous variations invented to investigate the
foundations of statistical mechanics, it is analyzable in full detail, see, \eg, \cite{CKDR11,CR12}, and can be
adapted to associate diffusion of points with further nonequilibrium thermodynamic properties and their
fluctuation theorems \cite{RTV00}.

Here, we engraft the Baker Map into the Slicer Map, giving each member a second coordinate $y$, uniform on $[0,1]$
and untouched during the flight. 
More precisely, we enlarge the phase space of the Slicer Map by introducing a new phase space $[0,2)\times [0,1]\times \mathbb{Z}$ whose coordinates are $x$, $y$, and $m$. The dynamics proceeds as follows. For the points belonging to the initial ensemble, namely those with $x\in [0,1)$ and $m=0$, the Slicer Map acts until a slicing event occurs. 
At that point, the $x$-coordinate of the moving element is transformed into $x + 1$: the member is thereby
marked as dead, once and for all, and its $x$ remains in $[1,2)$, evolving under the contracting half of the
Baker Map (Appendix E) with no effect on the displacement.
During this first stage of the dynamics, the coordinate $y$, which is initially uniformly distributed over $[0,1]$, does not change. In the second stage, that is, after the slicing event has occurred, the coordinates $y$ and, correspondingly, $m$, evolve according to the following rule:

\begin{align} \label{eq:bakerLift}
  y_{n+1} = 2 y_n \bmod 1 \, ,
  \qquad
  m_{n+1} = m_n + 2 \lfloor 2 y_n \rfloor - 1 \, ,
\end{align}
one cell per step, to the right or to the left according to the leading binary digit of $y_n$. Since $y$ is
independent of $x$ and its binary digits can be seen as the result of independent tosses of a fair coin, the
post-death motion turns out to be a simple random
walk with $D = 1/2$, in units of cells per step, with binomial propagators differing from the Gaussian ones at
relative order $1/n$ only; the dynamics is completed by the contracting half of the Baker Map, \appref{bakerDiff}.
The flight phase has zero Lyapunov exponent, the diffusive
phase $\ln 2$: the two extremes of chaos share the same chain of cells. The MSD obeys the exact identity
\begin{align} \label{eq:FaDmsd}
  \left\langle r^2 \right\rangle_t
  = \left\langle r^2 \right\rangle_t^{\mathrm{SM}}
  + 2 D \int_0^t \bigl[ 1 - W(s) \bigr] \, \rmd s \, ,
\end{align}
the frozen record of the Slicer Map plus the fresh diffusion accumulated since the end of each member's flight; the higher
moments follow from a single quadrature over the life-time distribution.

\begin{figure}[H]
  \includegraphics[width=\linewidth]{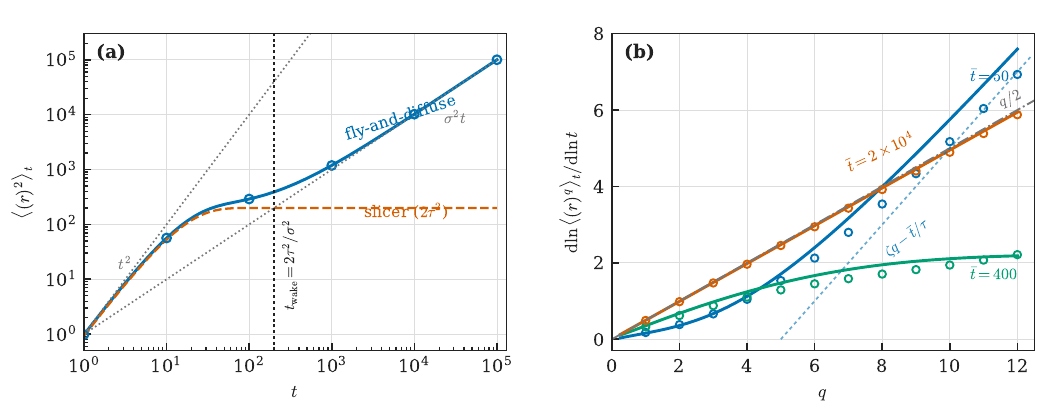}
  \caption{The fly-and-diffuse model with $W = \mathrm{e}^{-t/\tau}$, $\tau = 10$, and $D = 1/2$ ($\sigma^2 = 1$): symbols
  are simulations ($3 \times 10^5$ trajectories), lines the exact quadrature. (\textbf{a})~MSD: ballistic growth, the plateau
  $2\tau^2$ of the frozen flight record (the dashed line shows the Slicer Map with the same $W$, which localizes there), and
  the diffusive revival $\sigma^2 t$ beyond $t_{\rm wake} = 2\tau^2/\sigma^2$ (dotted vertical line). (\textbf{b})~Apparent
  spectra: at $\bar t = 50$ the bi-linear imitation, with the flight branch $\zeta\moment - \bar t/\tau$ (dotted); at
  $\bar t = 400$ the intermediate regime, in which the record is being washed out; at $\bar t = 2 \times 10^{4}$ the
  single-linear spectrum $\moment/2$ (dash-dotted) of normal diffusion: the gap between the slicer class and the Gaussian
  minimal model \eq{GBmodel} is closed.}
  \label{fig:FaD}
\end{figure}

Three regimes result (\figref{FaD}). For $t \lesssim \tau$ the motion is ballistic. For
$\tau \lesssim t \lesssim t_{\rm wake} := 2\tau^2/\sigma^2$, with $\sigma^2 = 2D$, the frozen flight record dominates the
bulk, and the model displays the slicer phenomenology of \secref{smSurvival}: an apparently bi-linear spectrum with branches
close to $0$ and $\zeta\moment - t/\tau$, and a knee drifting as $\varepsilon(t) = t/\tau$. For $t \gg t_{\rm wake}$,
diffusion has washed the record out, the PDF converges to the minimal model \eq{GBmodel}, and the MSD
resumes growing, $\langle r^2 \rangle_t \simeq 2\tau^2 + \sigma^2 (t - \tau)$. With the parameters of \figref{FaD}, the
flight branch has by then migrated to unobservably high moments, and the measured spectrum is single-linear,
$\zg_\moment \equiv 1$: the imitation of strong anomalous diffusion has died out, and what remains is normal transport,
carried by the bulk. A single particle model thus traverses the entire scenario of \secref{transient}, from the transient
bi-scaling of \figref{GaussTransient} to its resolution.

\subsection{Power-Law Flights over a Diffusive Bulk: the Fixed Knee}
\label{sec:gaussSAD}

The rule of \secref{crossRule} is realized exactly within the family. It suffices to revert the survival
function of the fly-and-diffuse model to the power law of \secref{SM}: the slicer sequence
\begin{align*}
  \ell_{|j|} = \frac{1}{2} \left( 1 + \frac{|j|}{t_0} \right)^{\!-p_2}
  \quad \text{yields} \quad
  W(t) = \left( 1 + \frac{t}{t_0} \right)^{\!-p_2} ,
\end{align*}
exactly at integer times, and the life-time function $t_c(x_0) = t_0\, (x_0^{-1/p_2} - 1)$ realizes the same $W$ at
all continuous times in the fly-and-die form. With $p_2 > 1$ the mean flight time is finite and the late-time bulk is
diffusive: $\theta = 1/2$, with the Gaussian scaling function \eq{GaussF}, $\sigma^2 = 2D$. To leading
order the moments are the sum of the Gaussian bulk, \eqn{GaussMoments}, and of the frozen flight record of the Slicer
Map, \eqn{SMgenM}:
\begin{align}\label{eq:gaussSADmoments}
  \left\langle r^\moment \right\rangle_t
  \simeq N_\moment\, t^{\moment/2} + \moment \int_0^t s^{\moment-1}\, W(s)\, \rmd s
  \simeq N_\moment\, t^{\moment/2} + \frac{\moment\, t_0^{p_2}}{\moment - p_2}\; t^{\moment - p_2}
  \qquad \text{for } \moment > p_2 \, , \; t \gg t_0 \, ,
\end{align}
The record integral is evaluated by the rescaling $s = u t$ of \appref{slicerW}: for $\moment > p_2$ it is
dominated by $u = O(1)$, where the shift in $W$ matters only to relative order $t_0/t$, and the $u$-integral
converges at $u = 0$ by the usual criterion; for $\moment < p_2$ the region $u \approx 0$ dominates instead, and the
record only adds a constant. Both branches are now time independent: this is
time-persistent strong anomalous diffusion, not the transient imitation of \secref{transient}, and the spectrum
reproduces \eq{crossRule} with $\zb = p_2$:
\begin{align}\label{eq:gaussSADspectrum}
  \frac{\moment\, \zg_\moment}{2}
  = \max\!\left( \frac{\moment}{2} \, , \; \moment - p_2 \right) ,
  \qquad
  \moment_c = \frac{\zb}{\zeta - \theta} = 2\, p_2 \, ,
\end{align}
with $\zeta = 1$ and $\zb = p_2$: the MSD is normal for every $p_2 > 1$, while all moments beyond $\moment_c = 2 p_2$
are flight dominated.

The distribution cannot have the matched structure of \eqn{ProbScalingCutoff}. Near the front the tail branch
is a clean power law, $\mathcal J(y) \simeq p_2\, t_0^{p_2}\, y^{-1-p_2}$ with $g(t) = t^{-\zb}$ and $\zb = p_2$,
extending down to $y \sim t^{\theta-\zeta}$; but under bulk rescaling the corresponding large-$z$ tail of
$t^{\theta}\,\mathcal P$ decays as $t^{-p_2/2}$. Clearly, the Gaussian scaling function does not develop a persistent power-law
tail, and the two branches do not overlap on a common intermediate asymptotics.

In the data analysis of \appref{parameterFitting}, the collapse of $\mathcal J$ with the fixed exponent
$\zb$ is perfect, but the joint bulk--tail plot of \figref{collapseTest}a cannot be built, because the bulk has no
power-law tail to match; the corrections to scaling of \appref{fitProblems} apply verbatim with
$\za \to \moment_c$.
We verified \eqns{gaussSADmoments} and \eq{gaussSADspectrum} by
exact sampling of the model ($2 \times 10^6$ trajectories, $p_2 = 3/2$, $t_0 = 1$, $\sigma = 1$): over the
tail-resolved fit window of \secref{SDCA}, the fitted slopes for $\moment = 1, \dots, 6$ are
$\{0.47,\, 0.93,\, 1.49,\, 2.42,\, 3.45,\, 4.47\}$, against
$\max(\moment/2, \moment - p_2) = \{0.5,\, 1,\, 1.5,\, 2.5,\, 3.5,\, 4.5\}$, with the largest deviations adjacent to
the crossover $\moment_c = 3$, as the corrections of \appref{fitProblems} prescribe; the amplitudes of
\eqn{gaussSADmoments} are matched within a few percent.

The renewal analogue is the L\'evy walk with flight time sampled from a Lévy distribution with exponent $p_3 > 2$: its bulk is Gaussian and diffusive ($\theta=1/2$), while the tail of the distribution is characterized by $\zeta=1$ and $\beta=p_3-1$, $\alpha=p_3$ so that $\moment_c =2\,(p_3 - 1)$, as
reported in the literature \cite{CMMGA99,ZDK15,VBB19,VBB20,ENBVB26}. Therefore, according to Corollary 1, the bulk and the tail are separated by a region where ${\cal {P}}$ is negligible.

\section{Conclusions}
\label{sec:conclusion}

We have employed methods of asymptotic matching, widely used in the context of hydrodynamic boundary layers and other scaling
theories, to deal with the large-distance cutoff of the PDF of the displacement in strong anomalous diffusion. We established
that strong anomalous transport emerges whenever the PDF features algebraic tails; conversely, in \appref{fastCutoff} we
prove that strong anomalous diffusion only arises if the distribution has power-law, or more slowly decaying, tails.

PDFs giving rise to strong anomalous diffusion are characterized by three exponents: $\theta$ describes how the width of the
distribution broadens in time, $\zeta$ describes the scaling of the cutoff, and $\za + 1$ is the exponent of the power-law
tails. The scaling of the moments of displacement is fully characterized by hyper-scaling relations based solely on the
moment order $\moment$ and these exponents. \eqn{momentsQ} relates the time dependence of the moments to $\moment$, $\za$,
$\zeta$, and $\theta$; this prediction rests purely on the constraints arising from matching the bulk and the tail of the
distribution. The slope $\theta$ of the lower part of the spectrum reflects the scaling of the bulk of the distribution, which
in most applications is entailed by large-deviation theory; the slope $\zeta$ of the upper part reflects the scaling of the
cutoff, which always arises for processes with bounded speed. The crossover between the two regimes is set by the exponent
$\za$ characterizing the power-law tail of the PDF, and the moment $\moment = \za$ grows as $t^{\za\theta} \ln t$,
cf.~\eqn{momentAlpha}---a result relevant, for instance, to the $t \ln t$ scaling of the MSD in the Lorentz gas with infinite
horizon \cite{CGLS14b,MT16,ZPFDB18}. The exponent $\zb$ describing the time scaling of the tail distribution is expressed in
terms of the other exponents by the hyper-scaling relation $\zb = \za\,(\zeta - \theta)$, \eqn{betadef}. Finally,
\eqn{momentScaling} provides the leading-order corrections to the power-law growth of the moments, and \eqn{critTimeScale}
the time scale that must be surpassed to clearly discriminate the leading-order power law from its sub-leading corrections.

The slowly decaying corrections to the power-law growth of the moments often give rise to severe systematic errors in the
estimated exponents, as demonstrated in \appref{fitProblems}. An informed way to determine $\za$, $\zeta$, and $\theta$, based
on the simultaneous scaling analysis of the bulk and of the tails of the PDF, was described in \appref{parameterFitting}.
Whenever the crossover time \eq{critTimeScale} exceeds the available observation time, fits of the moments are performed in
the sub-leading regime, and the resulting estimates of $\za$ can deviate substantially from the true value; the drifting
apparent exponents of \secref{transient} are the extreme form of this phenomenon.

We further determined what changes when the distribution is Gaussian, or, more generally, has no power-law tails. In that
case the spectrum of moments is single-linear ($\za = \infty$), and we proved (Proposition~\ref{prop:rigidity}) that this is
no accident: asymptotic matching between two different self-similar branches forces the intermediate asymptotics to be a
power law, so that power-law tails are necessary, and not only sufficient, for a time-persistent bi-linear spectrum. When
ballistic excursions are exponentially rare rather than power-law distributed, they produce only a transient imitation of
strong anomalous diffusion, which exists provided their persistence time exceeds the diffusive scale of the background
($\tau > 2\sigma^2$), and which even satisfies the hyper-scaling relation at any fixed observation time. The Gaussian
background is the marginal case: for heavier, stretched-exponential bulk tails the imitation even terminates at a finite time
(Remark~\ref{rem:stretched}). An exactly solvable deterministic realization is provided by the Slicer Map itself, once its
slicer sequence decays faster than any power law: the offset of the apparent second branch of the spectrum is then the
running exponent of the survival function (\secref{smSurvival}). Letting the members diffuse, rather than stop, at the end of
their flight yields the fly-and-diffuse model, which interpolates between the two scenarios and realizes the Gaussian minimal
model as a particle system (\secref{fad}). Its fingerprints are an apparent crossover moment that drifts linearly with the
observation time, crossover times growing linearly with the moment order, and the impossibility of collapsing the tails of
the distribution with any fixed exponent $\zb$.
Conversely, power-law flights over a Gaussian bulk do produce time-persistent strong anomalous
diffusion, even though the two branches of the distribution never match: the crossover moment is then set by the
crossing of the branches, $\moment_c = \zb/(\zeta - \theta)$---twice the tail exponent, for a diffusive bulk---rather
than by the tail exponent itself, and the logarithmic enhancement at the crossover is absent (\secref{gaussSAD}).

The hyper-scaling relations have been verified analytically for the Slicer Map and numerically for the spatial diffusion of
cold atoms---where the non-ballistic cutoff exponent $\zeta = 3/2$ puts the full three-exponent fitting scheme to work---and
for L\'evy walks, and they are consistent with the exponents reported in the literature for the systems collected in
\tabref{Spectrum}. With respect to Ref.~\cite{arxiv}, which is summarized in \secref{moments} and in \appref{fitProblems} and
\appref{parameterFitting}, the novel contributions of the present work are the complete analysis of the SDCA, based on newly
generated data and on the finite-sampling criterion for tail-dominated moments, and the converse theory of \secref{gaussian},
with its exactly solvable deterministic realizations. We conclude that our approach provides a unified perspective on the
analysis of the moments of displacement in strong anomalous diffusion, based on hyper-scaling relations and on a thorough
analysis of the time scaling of the PDF of the displacement. We expect it to be a robust tool for the analysis of
experimental and numerical data whenever power-law fits of individual moments are unreliable.

\vspace{6pt}

\authorcontributions{Conceptualization, methodology and formal analysis; writing---original draft preparation, review and
editing. All authors have read and agreed to the published version of the manuscript.}

\funding{This research was performed under the auspices of the Italian National Group of Mathematical Physics (GNFM) of
INdAM.}

\dataavailability{The numerical data presented in this study are available on request from the corresponding author.}

\conflictsofinterest{The authors declare no conflicts of interest.}

\newpage

\appendixtitles{yes}
\appendixstart
\appendix

\section{Corrections to Scaling}
\label{app:fitProblems}

This appendix recalls, from Ref.~\cite{arxiv}, the leading-order corrections to the asymptotic scaling of the moments, and
the resulting limits on power-law fits of individual moments.

\eqn{momentAlpha} establishes that there are logarithmic corrections to the power-law behavior for $\moment = \za$. The
resulting very weak time dependence features prominently in the discussion of the moments of displacement for the Lorentz gas
with infinite horizon \cite{B92,AHO03,CGLS14b,CESZ08,arxiv} and of the asymptotic scaling of its PDF \cite{D12,ZPFDB18}.

To assess the systematic errors made when fitting exponents for $\moment \neq \za$, we observe that \eqn{momentQ} predicts
that the \qmom{} moment of displacement is always a superposition of two power laws. The leading-order correction to the
asymptotic power-law scaling involves the ratio of the two prefactors,
\begin{subequations}\label{eq:momentScaling}%
\begin{align}
  \frac{\left\langle r^\moment \right\rangle}{I_\moment \, t^{\moment\zeta-\zb}}
  & \simeq
  1 + \left( \frac{N_\moment}{I_\moment} \right) t^{(\za - \moment)(\zeta - \theta)}
  \qquad\text{for } \za < \moment \, ,
  \\[5pt]
  \frac{\left\langle r^\moment \right\rangle}{N_\moment \, t^{\moment\theta}}
  & \simeq
  1 + \left( \frac{I_\moment}{N_\moment} \right) t^{(\moment - \za)(\zeta - \theta)}
  \qquad\text{for } \za > \moment \, .
\end{align}%
\end{subequations}%
For a reliable estimate of the exponents we demand that the leading-order term be larger than the correction term by a factor
$\Phi$. This requires that the power-law fits be performed for times well beyond the critical time $t_\moment$ defined by
\begin{subequations} \label{eq:critTimeScale}
\begin{align}
  t_\moment &= \left( \Phi \, N_\moment / I_\moment \right)^\tau
  \qquad \text{for } \za < \moment \, ,
  \\[3pt]
  t_\moment &= \left( \Phi \, I_\moment / N_\moment \right)^\tau
  \qquad \text{for } \za > \moment \, ,
\end{align}
with
\begin{align} \label{eq:tauDef}
  \tau = \frac{1}{\lvert \moment - \za \rvert \, (\zeta-\theta)} \, .
\end{align}
\end{subequations}

The time scale $t_\moment$ is model specific, but it can easily become very large: by \eqn{critTimeScale} it grows without
bounds as $\moment \to \za$, and the exponent $\tau$ is large whenever $\zeta - \theta$ is small. For the SDCA of \secref{SDCA},
for instance, $\zeta - \theta = 3/2 - 1/p_6$ vanishes as $p_6 \to 2/3$, the boundary of the strong anomalous phase, and the
fitted exponents of \figref{identifyParameters}b scatter accordingly around the predicted collapse, most severely for the
moments closest to the crossover.

Whether the fitted exponents are biased upward or downward depends on the signs of $N_\moment$ and $I_\moment$, \ie{} on
whether $\mathcal F$ and $\mathcal J$ approach the matching power law from below or from above (cf.~\secref{momentScalings});
in either case the bias persists as long as the subdominant power law in \eqn{momentQ} is not substantially smaller than the
leading order, \ie{} up to the times \eq{critTimeScale}.

In general, estimates of the moments of displacement will be hard whenever the difference between $\theta$ and $\zeta$ is small,
or when $\moment$ is close to the crossover moment $\za$. In order to avoid these issues, the exponents $\zeta$, $\theta$, and
$\za$ are determined from a scaling analysis of the PDF of the displacement, as described in \appref{parameterFitting}.

\section{Data Analysis of the PDFs}
\label{app:parameterFitting}

This appendix recalls the scheme of Ref.~\cite{arxiv} for determining the three exponents directly from the scaling of the
PDF; \secref{SDCA} applies it to the SDCA.

In order to extract the scaling exponents directly from the scaling of the PDF, we note that for small arguments
$\mathcal{J}(y)$ takes exactly the same functional form as $\mathcal{F}(z)$ for large arguments (cf.~\eqns{ProbTailZ} and
\eq{ProbTailV}). In \figref{collapseTest}a this is demonstrated by showing the scaling plots of the bulk and of the tails of
the PDF in the same coordinate system, rather than plotting them separately as in \figref{PDFscaling}. One clearly sees that
the large-$z$ data of $\mathcal{F}(z) = t^\theta\, \mathcal{P}(r, t)$ lie on the same power law as the small-$y$ data of
$\mathcal{J}(y) = t^{\zeta+\zb}\, \mathcal{P}(r, t)$.

The three exponents characterizing the distribution can therefore be found by a data collapse of the time evolution of the
PDFs of the displacement:
\begin{itemize}
\item[$\theta$] is used to collapse the height of the bulk distribution $\mathcal{F}(z) = t^\theta \, \mathcal{P}(r, t)$ for small
$z = r/t^\theta$. This one-parameter fit provides a data collapse for the bulk and the power-law tail of the distribution. It is
robust because it involves a vertical shift of horizontal lines such that they all meet at the same level. For the data in
\figref{collapseTest} this provides the estimate $\theta = 0.845 \pm 0.006$, for a theoretically expected value of
$\theta = 1/p_6 = 6/7 \simeq 0.857$. The resulting data collapse is shown in \figref{PDFscaling}a.

\item[$\zeta$] is used to collapse the position of the cutoff value $y_{\text{max}}$ of the tail distribution
$\mathcal{J}(y) = t^{\zeta+\zb} \, \mathcal{P}(r, t)$ for the largest admissible values of $y$; the vertical position of the
data is adjusted by the choice of $\zb$. The fit of $\zeta$ is robust because it involves the horizontal shift of very sharp
peaks such that they all lie at the same position; or rather, the minima at the end of the algebraic decay should lie at the
same height. In many circumstances $\zeta$ need not be fitted, because it must be unity when the large-scale cutoff is
dominated by ballistic flights; the SDCA of \secref{SDCA} is a case in which $\zeta$ must be fitted, since the
fastest atoms have velocities that themselves diffuse, and $\zeta = 3/2$. One obtains a plot of the data as shown in
\figref{PDFscaling}b.

\item[$\za$] must provide the slope of the \emph{simultaneous} asymptotic scaling of \emph{both} the large-$z$ asymptotics of
$\mathcal{F}(z)$ and the small-$y$ asymptotics of $\mathcal{J}(y)$. According to \eqn{betadef} it is related to $\zb$ by the
hyper-scaling relation $\zb = \za\,(\zeta-\theta)$. A joint plot of both distributions is shown in \figref{collapseTest}a. The
quality of the fit is best judged via a reduced plot displaying the deviation from the suggested power law: the power-law
scaling of $\mathcal{F}$ applies for $z \gtrsim z_c = 2$ and that of $\mathcal{J}$ for $y \lesssim y_c = 0.25$, and a reduced
plot of the data in this range is shown in \figref{collapseTest}b. The exponent $\zb$ must be adjusted such that all data
points take the same value. The effect of varying $\za$ for fixed $\zeta$ and $\theta$ is demonstrated in
\figref{collapseTest}c,d.
\end{itemize}
One clearly sees that the \emph{simultaneous} matching of the slopes of \emph{both} line segments provides a significantly
better grip on the data than a mere power-law fit. In the next section we show that this data analysis provides a reliable
characterization of numerical data.

\section{
Light tails rule out strong anomalous diffusion}
\label{app:fastCutoff}

In this appendix we show that weak anomalous diffusion emerges when the distribution $\mathcal P(r, t)$ features scaling for
small values of $r/t^\theta$ and decays faster than any power for large $r$. Scaling of the bulk of the distribution implies
that there are positive real numbers $\iota > \theta > 0$ and $c > 0$ such that for large times
\begin{subequations}
\begin{align}\label{eq:ProbScalingWeak}
  \mathcal{P}(r, t) \, \rmd r
  = \mathcal{F}(z) \, \rmd z
  \qquad \text{for} \quad z = \frac{r}{t^\theta} < c\, t^{\iota-\theta} \, .
\end{align}
At the same time, a decay of the tail faster than any power implies that for every $\moment > 0$ there are a constant
$k_\moment$ and a time $t_\moment$ such that
\begin{align} \label{eq:fastDecay}
  \mathcal{P}(r, t) < k_\moment \, t^{\theta} \, r^{-\moment-2}
  \qquad \text{for} \quad t > t_\moment \, .
\end{align}
\end{subequations}

We consider the following lower bound, $L_\moment(t)$, for the \qmom{} moment of the distribution,
\begin{align*}
  L_\moment(t)
  = \int_0^{c\, t^\iota} \rmd r \; r^\moment \, \mathcal{P}(r, t)
  \leq \Bigl\langle r^\moment \Bigr\rangle_t
  = L_\moment(t) + D_\moment(t) \, ,
  \qquad
  D_\moment(t) := \int_{c\, t^\iota}^\infty \rmd r \; r^\moment \, \mathcal{P}(r, t) \, ,
\end{align*}
which applies because the integrand is positive. The time evolution of $L_\moment$ is governed by the scaling of the bulk of
the PDF,
\begin{align*}
  L_\moment(t) = t^{\moment\,\theta} \int_0^{c\, t^{\iota-\theta}} \rmd z \; z^\moment \, \mathcal{F}(z)
  \;\xrightarrow[t \to \infty]{}\;
  N_\moment \, t^{\moment\,\theta} \, ,
  \qquad
  N_\moment = \int_0^\infty \rmd z \; z^\moment \, \mathcal{F}(z) \, ,
\end{align*}
where $N_\moment$ is a finite positive number if the moment exists. Further, the fast decay \eq{fastDecay} implies an upper
bound on $D_\moment(t)$,
\begin{align*}
  D_\moment(t)
  < k_\moment \, t^{\theta} \int_{c\, t^{\iota}}^\infty \rmd r \; r^{-2}
  = \frac{k_\moment}{c} \, t^{\theta-\iota} \, ,
\end{align*}
which tends to zero for large times, since $\iota > \theta$. The upper and lower bounds hence converge in the long-time limit,
and the moment must scale in time in the same way as $L_\moment(t)$:
\begin{align*}
  \Bigl\langle r^\moment \Bigr\rangle_t
  = N_\moment \, t^{\moment\,\theta} \qquad \text{for} \quad t \gg t_\moment \, .
\end{align*}
For $\theta = 1/2$ the system features normal diffusion, and otherwise weak anomalous diffusion: no strong anomalous diffusion
can arise when the tails decay faster than any power law.

\section{From Slicer Sequences to Survival Functions}
\label{app:slicerW}

This appendix derives \eqns{SMgenP} and \eq{SMgenM} from the map \eq{SMdef}. Two structural facts make the derivation
elementary: the map never changes the coordinate $x$, and the slicer sequence decreases monotonically.

Consider the right-moving stream, $x \in (1/2, 1]$, and let $u := 1 - x \in [0, 1/2)$; since $\ell_0 = 1/2$, this whole
stream moves to cell $1$ at the first step. A member arriving in cell $j$---which happens at time $j$, one cell per
step---proceeds to cell $j+1$ if and only if $x > 1 - \ell_j$, \ie{} $u < \ell_j$; otherwise it is sent back, and it
then bounces between cells $j-1$ and $j$ forever, since it cleared the slicer of cell $j-1$ but not that of cell $j$.
A member is therefore still flying at time $m$ if and only if it has cleared every cell up to $m$:
\begin{align*}
  u < \ell_j \; \text{ for all } \; j \le m
  \qquad \Longleftrightarrow \qquad
  u < \ell_m \, ,
\end{align*}
where the equivalence holds by monotonicity: only the last, lowest slicer matters. With $x$ uniformly distributed, each
of the two symmetric streams carries a mass $\ell_m$ of survivors, so that the flying fraction is
\begin{align}\label{eq:WfromEll}
  W(m) = 2\, \ell_m \qquad \text{exactly, at all integer times;}
\end{align}
conversely, the sequence $\ell_{|j|} = W(j)/2$ realizes any non-increasing survival function with $W(0) = 1$. Each
member is thus characterized by a life time $t_c$ with $\operatorname{Prob}(t_c > s) = W(s)$---in the fly-and-die form,
$t_c(x_0) = W^{-1}(x_0)$ with $x_0$ uniform on $(0,1]$, the inverse-transform sampling of $W$---and its displacement
is, up to the bounded bouncing,
\begin{align}\label{eq:lifetimeRep}
  R(t) = \min\left( t_c \, , \; t \right) .
\end{align}
Everything follows from this representation.

To obtain \eqn{SMgenP}, let us define the ensemble average at time $t$ of an arbitrary observable $\varphi$ as its
integral over the label of the ensemble---the initial coordinate $x_0$, the only random element of the deterministic
dynamics:
\begin{align*}
  \left\langle \varphi(R(t)) \right\rangle
  := \int_0^1 \varphi\!\left( R(x_0, t) \right) \rmd x_0 \, ,
\end{align*}
where it is assumed that the initial distribution of points is uniform on $(0,1]$.
Changing variables from $x_0$ to the life time, $t_c = W^{-1}(x_0)$---monotone, with $\rmd x_0 = -\rmd W(t_c)$---and
splitting the members into those that are dead at time $t$, with $t_c \le t$ and frozen at $r = t_c$, and those that
are still flying, with $t_c > t$ and located at $r = t$, one finds
\begin{align*}
  \left\langle \varphi(R(t)) \right\rangle
  = \int_0^t \varphi(s) \left( -W'(s) \right) \rmd s + \varphi(t)\, W(t) \, ,
\end{align*}
since $-W'(s)\,\rmd s$ is the life-time probability element, $\operatorname{Prob}\!\left( t_c \in [s, s + \rmd s)
\right)$.
The integration variable $s$ has therefore three equivalent readings: it is the life time $t_c$ labelling the
members---the fundamental one, being the ensemble measure pushed forward from $x_0$; it is the frozen position $r$ of
the dead members, since $r = t_c$ at unit speed; and its range $(0, t)$ is the support of the frozen record.
As $\varphi$ is arbitrary, the distribution of $R(t)$ is
\begin{align*}
  \mathcal{P}(r,t) = -W'(r)\,\Theta(t-r) + W(t)\,\delta(r-t) \, ,
\end{align*}
which is \eqn{SMgenP}: the frozen record inherits the life-time density---position equals time of death---while the
flying cohort rides the front. Normalization is automatic, $[W(0) - W(t)] + W(t) = 1$. If $W$ is not differentiable,
$-W'(r)\,\rmd r$ is read as the Stieltjes measure $-\rmd W(r)$. In particular, for the map at integer times $W$ is a
step function, constant on the intervals $[k, k+1)$: its classical derivative vanishes almost everywhere, while
$-\rmd W$ is the purely atomic measure $\sum_k \left[ W(k-1) - W(k) \right] \delta(r - k)\, \rmd r$, so that
\eqn{SMgenP} becomes
\begin{align}\label{eq:atomicP}
  \mathcal{P}(r, n) = \sum_{k \le n} \left[ W(k-1) - W(k) \right] \delta(r - k) + W(n)\, \delta(r - n) \, ,
\end{align}
the exact distribution of the frozen cells, whose weights are twice the strip lengths $\Delta_k$ of
Ref.~\cite{GRTV17}. The smooth form of \eqn{SMgenP} describes the equivalent continuous-time fly-and-die process, of
which the map is the integer-time sampling,
and the difference between the two descriptions affects the moments only at relative order $1/t$.

To obtain \eqn{SMgenM}, multiply \eqn{SMgenP} by $r^\moment$ and integrate: the $\Theta$ term restricts the
integration to the frozen record, the $\delta$ term evaluates it at the front, and
\begin{align}\label{eq:momAverage}
  \left\langle r^\moment \right\rangle_t
  = \int_0^\infty r^\moment \, \mathcal{P}(r,t)\, \rmd r
  = \int_0^t s^\moment \left( -W'(s) \right) \rmd s + t^\moment\, W(t) \, ,
\end{align}
which depends on $t$ in three places: the upper limit of the integral, and both factors of the flying term.
Differentiating, we get:
\begin{align*}
  \frac{\rmd}{\rmd t} \left\langle r^\moment \right\rangle_t
  = - t^\moment\, W'(t) + \moment\, t^{\moment-1}\, W(t) + t^\moment\, W'(t)
  = \moment\, t^{\moment-1}\, W(t) \, .
\end{align*}
In the interval $\rmd t$ the dying mass $-W'(t)\,\rmd t$ leaves the front and is deposited in the record at the very
same position, $r = t$, so that death, by itself, changes no moment; the only net effect is the ballistic transport of
the still-flying mass. Integration from $R(0) = 0$ then gives the compact form of \eqn{SMgenM}, and the logarithmic
slope \eq{SMgenMln} is its immediate consequence.

To obtain \eqn{SMgenBranches}, rescale $s = u t$ in the compact form of \eqn{SMgenM}, so that
\begin{align*}
  \left\langle r^\moment \right\rangle_t
  = \moment \int_0^t s^{\moment - 1}\, W(s)\, \rmd s
  = \moment\, t^{\moment} \int_0^1 u^{\moment - 1}\, W(u t)\, \rmd u \, ,
\end{align*}
and substitute it into \eq{SMgenMln}: the prefactor $\moment\, t^{\moment}$ cancels between numerator and denominator,
leaving
\begin{align*}
  \frac{\rmd \ln \left\langle r^\moment \right\rangle_t}{\rmd \ln t}
  = \frac{\rmd \ln \left\langle r^\moment \right\rangle_t}{\rmd t}\, \frac{\rmd t}{\rmd \ln t}
  = \frac{\moment\, t^{\moment}\, W(t)}{\moment\, t^{\moment} \int_0^1 u^{\moment - 1}\, W(u t)\, \rmd u}
  = \left[ \, \int_0^1 u^{\moment - 1}\, \frac{W(u t)}{W(t)}\, \rmd u \right]^{-1} .
\end{align*}
where we have:
\begin{align*}
  \int_{u t}^{t} \varepsilon(s)\, \rmd \ln s
  = - \int_{u t}^{t} \rmd \ln W
  = \ln \frac{W(u t)}{W(t)} \, .
\end{align*}
Therefore, wherever $\varepsilon$ varies little over logarithmic time scales, we can write
$\varepsilon(s) \simeq \varepsilon(t)$ for $s \in (ut, t)$, and $W(u t)/W(t) \simeq u^{-\varepsilon(t)}$, which
implies:
\begin{align*}
  \frac{\rmd \ln \left\langle r^\moment \right\rangle_t}{\rmd \ln t}
  \simeq \left[ \, \int_0^1 u^{\moment - 1 - \varepsilon(t)}\, \rmd u \right]^{-1}
  = \left[ \frac{1}{\moment - \varepsilon(t)} \right]^{-1}
  = \moment - \varepsilon(t)
  \qquad \text{for} \quad \moment > \varepsilon(t) \, ,
\end{align*}
The $u$-integral converges at $u = 0$ if and only if $\moment > \varepsilon(t)$: this is the origin of the restriction
in the preceding display. The result is the upper branch of \eqn{SMgenBranches}, with $\zeta = 1$. For the pure power
law, we have $W \propto t^{-p_2}$, hence $\varepsilon \equiv p_2$, and the ratio $W(u t)/W(t) = u^{-p_2}$ holds
identically, so that the upper branch is exact at all times.

For $\moment < \varepsilon(t)$, instead, the $u$-integral draws its mass from the neighborhood of $u = 0$---small $u$
corresponds to early times, $s = u t$---and the moment stops growing: $\left\langle r^\moment \right\rangle_t$
saturates to the finite limit
$\int_0^\infty r^\moment \left( -\rmd W(r) \right)$, the \qmom{} moment
of the life time, since every member is eventually frozen at $r = t_c$.
Meanwhile $\varepsilon(t)$ exceeds $\moment$ from some time on, so the numerator
$\moment\, t^{\moment}\, W(t)$ of \eq{SMgenMln} vanishes as $t$ grows; equivalently, the rescaled $u$-integral grows
without bound, and its inverse---the logarithmic slope---tends to zero. This is the lower branch of
\eqn{SMgenBranches}. The knee sits where the two behaviors exchange dominance, at $\moment \approx \varepsilon(t)$.

An independent route to \eqn{SMgenM} is the one that generalizes
beyond single-time quantities as in Ref.\cite{GRTV17}:
writing $R(t) = \int_0^t \Theta(t_c - s)\, \rmd s$ and raising it to the \qmom{} power,
\begin{align*}
  \left\langle r^\moment \right\rangle_t
  = \int_{[0,t]^\moment} W\!\left( \max_i s_i \right) \rmd^\moment s
  = \moment \int_0^t u^{\moment - 1}\, W(u)\, \rmd u \, ,
\end{align*}
since $\langle \Theta(t_c - s_1) \cdots \Theta(t_c - s_\moment) \rangle = \operatorname{Prob}(t_c > \max_i s_i) = W(\max_i s_i)$, the average being the $x_0$-integral
defined above, and the shell
$\max_i s_i \in (u, u + \rmd u)$ has volume $\rmd (u^\moment)$.
The same representation extends to multi-time
quantities; for instance, $\langle R(m)\, R(n) \rangle = \int_0^m \!\! \int_0^n W\!\left( \max(s, s') \right)
\rmd s \, \rmd s'$, whose power-law case is studied in Ref.~\cite{GRTV17}.

Note also that \eqn{SMgenM} involves only $W$ itself, never its derivative, and that its compact form is exact at
integer times for the step function as well: on each unit cell,
\begin{align*}
  \moment \int_k^{k+1} s^{\moment - 1}\, W(s)\, \rmd s
  = W(k) \left[ (k+1)^\moment - k^\moment \right] ,
\end{align*}
and the sum over the cells reproduces, upon rearrangement of the terms, the moments of the atomic distribution
\eq{atomicP}, $\langle r^\moment \rangle_n = \sum_{k \le n} k^\moment \left[ W(k-1) - W(k) \right] + n^\moment\, W(n)$.
Indeed, the one-step increment of the latter mirrors the continuous-time cancellation above:
\begin{align*}
  \left\langle r^\moment \right\rangle_{n+1} - \left\langle r^\moment \right\rangle_n
  &= (n+1)^\moment \left[ W(n) - W(n+1) \right] + (n+1)^\moment\, W(n+1) - n^\moment\, W(n)
  \\
  &= \left[ (n+1)^\moment - n^\moment \right] W(n) \, ,
\end{align*}
the discrete twin of $\moment\, t^{\moment-1}\, W(t)\, \rmd t$: the set of points dying during the step appears in the record
at $(n+1)^\moment$ and cancels against its disappearance from the front, leaving only the ballistic advance of the mass
$W(n)$ that flies through the step.

\section{Deterministic Diffusion in the Fly-and-Diffuse Model}
\label{app:bakerDiff}

This appendix spells out the deterministic chaotic version of the fly-and-diffuse model of \secref{fad} and derives
its exact displacement statistics. The phase space is $[0,2) \times [0,1] \times \mathbb{Z}$: a member carries the Slicer
coordinate $x$, the Baker coordinate $y$, and its cell $m$. Flying members, $x \in [0,1)$, follow the Slicer
Map \eq{SMdef}, advancing one cell per step with $y$ untouched; at the slicing event they are marked once and for
all by $x \mapsto x + 1$. Dead members, $x \in [1,2)$, evolve by
\begin{align}\label{eq:bakerFull}
  x' = 1 + \frac{x - 1 + \lfloor 2y \rfloor}{2} \, ,
  \qquad
  y' = 2y \bmod 1 \, ,
  \qquad
  m' = m + 2 \lfloor 2y \rfloor - 1 \, ,
\end{align}
the Baker transformation of the unit square, translated to the strip $x \in [1,2)$ and extended by one cell jump per
step: it is invertible, it preserves the Lebesgue measure, and its stretching rate gives the Lyapunov exponent
$\ln 2$, against the exponent zero of the flight phase, which is a translation. The first equation of
\eq{bakerFull} plays no role in the displacement: the pair $(y, m)$ evolves autonomously, and every
statistic below involves $y$ alone. It is what makes the dead-phase dynamics invertible---exponents
$\pm \ln 2$, as in a volume-preserving system---while the mark $x \mapsto x + 1$ records death as part
of the state, keeping the whole evolution a single map on the phase space.

The jumps are governed by the binary digits of $y$. Writing $y = \sum_{i \ge 1} b_i\, 2^{-i}$ with $b_i \in \{0,1\}$,
the doubling $y \mapsto 2y \bmod 1$ deletes the leading digit, so that the $i$-th diffusive step moves the member by
$\sigma_i = 2 b_i - 1 = \pm 1$. With $y$ uniformly distributed, the digits are independent fair coins: the event
$\{ b_1 = \beta_1, \dots, b_j = \beta_j \}$ is a dyadic interval of length $2^{-j}$, whatever the $\beta_i$.
In the language of symbolic dynamics, the digit string is the itinerary of $y$ over the partition
$\{[0,1/2), [1/2,1)\}$, on which the doubling acts as the shift, and the Lebesgue measure reads as the Bernoulli
$(1/2,1/2)$ measure on the symbols: the deterministic map and the coin-tossing process become
probabilistically---\ie{} in a coarse-grained sense---the same object under two names. We have thus introduced a
diffusive phase as a simple random walk,
\begin{align}\label{eq:binomProp}
  \operatorname{Prob}\left( \sigma_1 + \dots + \sigma_j = \ell \right)
  = \Pi_j(\ell)
  = 2^{-j} \binom{j}{(j+\ell)/2}
  \quad \text{for } j + \ell \text{ even} \text{ and } \ell \in \{-j,\ldots  ,j \} 
\end{align}
with zero mean and variance $j$: in the notation of \secref{fad}, $\sigma^2 = 1$, \ie{} $D = 1/2$, in units of cells
per step.

Independence completes the solution. The life time depends on $x$ alone, the walk on $y$ alone, and the initial
measure is a product, so the displacement law of the right-moving stream is the exact convolution
\begin{align}\label{eq:bakerConv}
  \operatorname{Prob}\left( m_n = r \right)
  = \sum_{k \le n} \left[ W(k-1) - W(k) \right] \Pi_{n-k}(r - k) + W(n)\, \delta_{r, n} \, ,
\end{align}
the life-time atoms of \eq{atomicP} smeared by the walk accumulated since each death; the left-moving stream is its
mirror image. 
The computation of the second moment
\begin{align*}
  \langle m_n^2 \rangle = \sum_{k \le n} \left[ W(k-1) - W(k) \right] \left[ k^2 + (n-k) \right] + n^2\, W(n) \, ,
\end{align*}
together with
\begin{align*}
  \sum_{k \le n} \left[ W(k-1) - W(k) \right] (n-k) = \sum_{i=1}^{n-1} \left[ 1 - W(i) \right]
\end{align*}
shows that the fly-and-diffuse identity \eq{FaDmsd} holds exactly at integer times, with $2D = 1$. By the
de~Moivre--Laplace theorem for binomial distributions,
$\Pi_j$ converges to the Gaussian kernel of variance $j$ at relative order $1/j$, so all
the formulas of \secref{fad} are recovered. We verified \eq{bakerConv} and the MSD by iterating the map in exact
dyadic arithmetic ($2 \times 10^5$ members, $\tau = 10$, $n = 200$), with agreement to statistical accuracy.

\vspace{1cm}

\title


{\large References}
\vspace{1cm}

\begin{thebibliography}{99}

\bibitem{Metzler}
Metzler, R.; Klafter, J. The random walk's guide to anomalous diffusion: A fractional dynamics approach. \emph{Phys.\ Rep.}
\textbf{2000}, \emph{339}, 1. \url{https://doi.org/10.1016/S0370-1573(00)00070-3}

\bibitem{KRS08}
Klages, R.; Radons, G.; Sokolov, I.M. (Eds.) \emph{Anomalous Transport: Foundations and Applications}; Wiley-VCH: Weinheim,
Germany, 2008. \url{https://doi.org/10.1002/9783527622979}

\bibitem{Sok12}
Sokolov, I.M. Models of anomalous diffusion in crowded environments. \emph{Soft Matter} \textbf{2012}, \emph{8}, 9043.
\url{https://doi.org/10.1039/C2SM25701G}

\bibitem{frontiers}
Mej\'{i}a-Monasterio, C.; Metzler, R.; Vollmer, J. Anomalous transport: Applications, mathematical perspectives, and big
data. \emph{Front.\ Phys.} \textbf{2020}, \emph{8}, 622417. \url{https://doi.org/10.3389/fphy.2020.622417}

\bibitem{Saxton}
Saxton, M.J. Anomalous subdiffusion in fluorescence photobleaching recovery: A Monte Carlo study. \emph{Biophys.\ J.}
\textbf{2001}, \emph{81}, 2226. \url{https://doi.org/10.1016/S0006-3495(01)75870-5}

\bibitem{GC06}
Golding, I.; Cox, E.C. Physical nature of bacterial cytoplasm. \emph{Phys.\ Rev.\ Lett.} \textbf{2006}, \emph{96}, 098102.
\url{https://doi.org/10.1103/PhysRevLett.96.098102}

\bibitem{NHB07}
Nicolau, D.V.; Hancock, J.F.; Burrage, K. Sources of anomalous diffusion on cell membranes: A Monte Carlo study.
\emph{Biophys.\ J.} \textbf{2007}, \emph{92}, 1975. \url{https://doi.org/10.1529/biophysj.105.076869}

\bibitem{Rina}
Martin, R.L.; Jerolmack, D.J.; Schumer, R. The physical basis for anomalous diffusion in bed load transport. \emph{J.\
Geophys.\ Res.\ Earth Surf.} \textbf{2012}, \emph{117}, F01018. \url{https://doi.org/10.1029/2011JF002075}

\bibitem{LLP}
Lepri, S.; Livi, R.; Politi, A. Thermal conduction in classical low-dimensional lattices. \emph{Phys.\ Rep.} \textbf{2003},
\emph{377}, 1. \url{https://doi.org/10.1016/S0370-1573(02)00558-6}

\bibitem{AHO03}
Armstead, D.N.; Hunt, B.R.; Ott, E. Anomalous diffusion in infinite horizon billiards. \emph{Phys.\ Rev.\ E} \textbf{2003},
\emph{67}, 021110. \url{https://doi.org/10.1103/PhysRevE.67.021110}

\bibitem{JBR08}
Jepps, O.G.; Bianca, C.; Rondoni, L. Onset of diffusive behavior in confined transport systems. \emph{Chaos} \textbf{2008},
\emph{18}, 013127. \url{https://doi.org/10.1063/1.2888759}

\bibitem{CMMGA99}
Castiglione, P.; Mazzino, A.; Muratore-Ginanneschi, P.; Vulpiani, A. On strong anomalous diffusion. \emph{Physica D}
\textbf{1999}, \emph{134}, 75. \url{https://doi.org/10.1016/S0167-2789(99)00031-7}

\bibitem{VRTGM21}
Vollmer, J.; Rondoni, L.; Tayyab, M.; Giberti, C.; Mej\'{i}a-Monasterio, C. Displacement autocorrelation functions for strong
anomalous diffusion: A scaling form, universal behavior, and corrections to scaling. \emph{Phys.\ Rev.\ Res.} \textbf{2021},
\emph{3}, 013067. \url{https://doi.org/10.1103/PhysRevResearch.3.013067}

\bibitem{SL06}
Sanders, D.P.; Larralde, H. Occurrence of normal and anomalous diffusion in polygonal billiard channels. \emph{Phys.\ Rev.\
E} \textbf{2006}, \emph{73}, 026205. \url{https://doi.org/10.1103/PhysRevE.73.026205}

\bibitem{JR06}
Jepps, O.G.; Rondoni, L. Thermodynamics and complexity of simple transport phenomena. \emph{J.\ Phys.\ A Math.\ Gen.}
\textbf{2006}, \emph{39}, 1311. \url{https://doi.org/10.1088/0305-4470/39/6/007}

\bibitem{Orchard21}
Orchard, J.; Rondoni, L.; Mej\'{i}a-Monasterio, C.; Frascoli, F. Diffusion and escape from polygonal channels: Extreme values
and geometric effects. \emph{J.\ Stat.\ Mech.\ Theory Exp.} \textbf{2021}, \emph{2021}, 073208.
\url{https://doi.org/10.1088/1742-5468/ac0c71}

\bibitem{SS06}
Schmiedeberg, M.; Stark, H. Superdiffusion in a honeycomb billiard. \emph{Phys.\ Rev.\ E} \textbf{2006}, \emph{73}, 031113.
\url{https://doi.org/10.1103/PhysRevE.73.031113}

\bibitem{CESZ08}
Courbage, M.; Edelman, M.; Fathi, S.M.S.; Zaslavsky, G.M. Problem of transport in billiards with infinite horizon.
\emph{Phys.\ Rev.\ E} \textbf{2008}, \emph{77}, 036203. \url{https://doi.org/10.1103/PhysRevE.77.036203}

\bibitem{AKB17}
Aghion, E.; Kessler, D.A.; Barkai, E. Large fluctuations for spatial diffusion of cold atoms. \emph{Phys.\ Rev.\ Lett.}
\textbf{2017}, \emph{118}, 260601. \url{https://doi.org/10.1103/PhysRevLett.118.260601}

\bibitem{Pik91}
Pikovsky, A.S. Statistical properties of dynamically generated anomalous diffusion. \emph{Phys.\ Rev.\ A} \textbf{1991},
\emph{43}, 3146. \url{https://doi.org/10.1103/PhysRevA.43.3146}

\bibitem{SRGK15}
Salari, L.; Rondoni, L.; Giberti, C.; Klages, R. A simple non-chaotic map generating subdiffusive, diffusive, and
superdiffusive dynamics. \emph{Chaos} \textbf{2015}, \emph{25}, 073113. \url{https://doi.org/10.1063/1.4926621}

\bibitem{AC03}
Artuso, R.; Cristadoro, G. Anomalous transport: A deterministic approach. \emph{Phys.\ Rev.\ Lett.} \textbf{2003}, \emph{90},
244101. \url{https://doi.org/10.1103/PhysRevLett.90.244101}

\bibitem{GW10}
Gal, N.; Weihs, D. Experimental evidence of strong anomalous diffusion in living cells. \emph{Phys.\ Rev.\ E} \textbf{2010},
\emph{81}, 020903(R). \url{https://doi.org/10.1103/PhysRevE.81.020903}

\bibitem{SJ94}
Sneppen, K.; Jensen, M.H. Multidiffusion in critical dynamics of strings and membranes. \emph{Phys.\ Rev.\ E} \textbf{1994},
\emph{49}, 919. \url{https://doi.org/10.1103/PhysRevE.49.919}

\bibitem{CLNZ99}
Carreras, B.A.; Lynch, V.E.; Newman, D.E.; Zaslavsky, G.M. Anomalous diffusion in running sandpile models. \emph{Phys.\ Rev.\
E} \textbf{1999}, \emph{60}, 4770. \url{https://doi.org/10.1103/PhysRevE.60.4770}

\bibitem{BCV10}
Burioni, R.; Caniparoli, L.; Vezzani, A. L\'evy walks and scaling in quenched disordered media. \emph{Phys.\ Rev.\ E}
\textbf{2010}, \emph{81}, 060101(R). \url{https://doi.org/10.1103/PhysRevE.81.060101}

\bibitem{BVLV14}
Bernab\'o, P.; Burioni, R.; Lepri, S.; Vezzani, A. Anomalous transmission and drifts in one-dimensional L\'evy structures.
\emph{Chaos Solitons Fractals} \textbf{2014}, \emph{67}, 11. \url{https://doi.org/10.1016/j.chaos.2014.06.002}

\bibitem{RDHB14a}
Rebenshtok, A.; Denisov, S.; H\"anggi, P.; Barkai, E. Non-normalizable densities in strong anomalous diffusion: Beyond the
central limit theorem. \emph{Phys.\ Rev.\ Lett.} \textbf{2014}, \emph{112}, 110601.
\url{https://doi.org/10.1103/PhysRevLett.112.110601}

\bibitem{RDHB14b}
Rebenshtok, A.; Denisov, S.; H\"anggi, P.; Barkai, E. Infinite densities for L\'evy walks. \emph{Phys.\ Rev.\ E}
\textbf{2014}, \emph{90}, 062135. \url{https://doi.org/10.1103/PhysRevE.90.062135}

\bibitem{FSBZ15}
Froemberg, D.; Schmiedeberg, M.; Barkai, E.; Zaburdaev, V. Asymptotic densities of ballistic L\'evy walks. \emph{Phys.\ Rev.\
E} \textbf{2015}, \emph{91}, 022131. \url{https://doi.org/10.1103/PhysRevE.91.022131}

\bibitem{ZDK15}
Zaburdaev, V.; Denisov, S.; Klafter, J. L\'evy walks. \emph{Rev.\ Mod.\ Phys.} \textbf{2015}, \emph{87}, 483.
\url{https://doi.org/10.1103/RevModPhys.87.483}


\bibitem{VBB19}
Vezzani, A.; Barkai, E.; Burioni, R. Single-big-jump principle in physical modeling. \emph{Phys.\ Rev.\ E} \textbf{2019}, \emph{100}, 012108.
\url{https://doi.org/10.1103/PhysRevE.100.012108}

\bibitem{VBB20}
Vezzani, A.; Barkai, E.; Burioni, R. Rare events in generalized Lévy Walks and the Big Jump principle. \emph{Sci. Rep.} \textbf{2020}, \emph{10}, 2732.
\url{https://doi.org/10.1038/s41598-020-59187-w}


\bibitem{ENBVB26}
Ellettari, E.; Nasuti, G.; Bassanoni, A.; Vezzani, A.; Burioni, R. Rare events and redundancy in random walkers target search in a finite domain. \emph{Chaos Solit. Fractals} \textbf{2026}, \emph{210}, 118650.
\url{https://doi.org/10.1016/j.chaos.2026.118650}

\bibitem{KCNP16}
Krapf, D.; Campagnola, G.; Nepal, K.; Peersen, O.B. Strange kinetics of bulk-mediated diffusion on lipid bilayers.
\emph{Phys.\ Chem.\ Chem.\ Phys.} \textbf{2016}, \emph{18}, 12633. \url{https://doi.org/10.1039/C6CP00937A}

\bibitem{Barenblatt}
Barenblatt, G.I. \emph{Scaling}; Cambridge University Press: Cambridge, UK, 2003.

\bibitem{BO}
Bender, C.M.; Orszag, S.A. \emph{Advanced Mathematical Methods for Scientists and Engineers}; McGraw-Hill: Singapore, 1978.

\bibitem{arxiv}
Vollmer, J.; Giberti, C.; Orchard, J.; Reinhard, H.; Mej\'{i}a-Monasterio, C.; Rondoni, L. Universal hyper-scaling relations,
power-law tails, and data analysis for strong anomalous diffusion. \emph{arXiv} \textbf{2024}, arXiv:2412.20590.
\url{https://doi.org/10.48550/arXiv.2412.20590}

\bibitem{MEZ96}
Marksteiner, S.; Ellinger, K.; Zoller, P. Anomalous diffusion and L\'evy walks in optical lattices. \emph{Phys.\ Rev.\ A}
\textbf{1996}, \emph{53}, 3409. \url{https://doi.org/10.1103/PhysRevA.53.3409}

\bibitem{KB12}
Kessler, D.A.; Barkai, E. Theory of fractional L\'evy kinetics for cold atoms diffusing in optical lattices. \emph{Phys.\
Rev.\ Lett.} \textbf{2012}, \emph{108}, 230602. \url{https://doi.org/10.1103/PhysRevLett.108.230602}

\bibitem{D14}
Dettmann, C.P. Diffusion in the Lorentz gas. \emph{Commun.\ Theor.\ Phys.} \textbf{2014}, \emph{62}, 521.
\url{https://doi.org/10.1088/0253-6102/62/4/10}

\bibitem{KB10}
Kessler, D.A.; Barkai, E. Infinite covariant density for diffusion in logarithmic potentials and optical lattices.
\emph{Phys.\ Rev.\ Lett.} \textbf{2010}, \emph{105}, 120602. \url{https://doi.org/10.1103/PhysRevLett.105.120602}

\bibitem{DL12}
Dechant, A.; Lutz, E. Anomalous spatial diffusion and multifractality in optical lattices. \emph{Phys.\ Rev.\ Lett.}
\textbf{2012}, \emph{108}, 230601. \url{https://doi.org/10.1103/PhysRevLett.108.230601}

\bibitem{GRTV17}
Giberti, C.; Rondoni, L.; Tayyab, M.; Vollmer, J. Equivalence of position-position auto-correlations in the Slicer
Map and the L\'evy--Lorentz gas.
\emph{Nonlinearity} \textbf{2019}, \emph{32}, 2302. \url{https://doi.org/10.1088/1361-6544/ab08f6}

\bibitem{1989BlumenZumofenKlafter}
Blumen, A.; Zumofen, G.; Klafter, J. Transport aspects in anomalous diffusion: L\'evy walks. \emph{Phys.\ Rev.\ A}
\textbf{1989}, \emph{40}, 3964. \url{https://doi.org/10.1103/PhysRevA.40.3964}

\bibitem{DBSB15}
Dentz, M.; Le~Borgne, T.; Lester, D.R.; de~Barros, F.P.J. Scaling forms of particle densities for L\'evy walks and strong
anomalous diffusion. \emph{Phys.\ Rev.\ E} \textbf{2015}, \emph{92}, 032128. \url{https://doi.org/10.1103/PhysRevE.92.032128}

\bibitem{MVN68}
Mandelbrot, B.B.; Van~Ness, J.W. Fractional Brownian motions, fractional noises and applications. \emph{SIAM Rev.}
\textbf{1968}, \emph{10}, 422. \url{https://doi.org/10.1137/1010093}

\bibitem{Goldstein51}
Goldstein, S. On diffusion by discontinuous movements, and on the telegraph equation. \emph{Q.\ J.\ Mech.\ Appl.\ Math.}
\textbf{1951}, \emph{4}, 129. \url{https://doi.org/10.1093/qjmam/4.2.129}

\bibitem{Kac74}
Kac, M. A stochastic model related to the telegrapher's equation. \emph{Rocky Mt.\ J.\ Math.} \textbf{1974}, \emph{4}, 497.
\url{https://doi.org/10.1216/RMJ-1974-4-3-497}

\bibitem{GN82}
Geisel, T.; Nierwetberg, J. Onset of diffusion and universal scaling in chaotic systems. \emph{Phys.\ Rev.\ Lett.}
\textbf{1982}, \emph{48}, 7. \url{https://doi.org/10.1103/PhysRevLett.48.7}

\bibitem{GF82}
Grossmann, S.; Fujisaka, H. Diffusion in discrete nonlinear dynamical systems. \emph{Phys.\ Rev.\ A}
\textbf{1982}, \emph{26}, 1779. \url{https://doi.org/10.1103/PhysRevA.26.1779}

\bibitem{Klages07}
Klages, R. \emph{Microscopic Chaos, Fractals and Transport in Nonequilibrium Statistical Mechanics}; Advanced Series
in Nonlinear Dynamics Vol.~24; World Scientific: Singapore, 2007. \url{https://doi.org/10.1142/5945}

\bibitem{CKDR11}
Colangeli, M.; Klages, R.; De~Gregorio, P.; Rondoni, L. Steady state fluctuation relations and time reversibility
for non-smooth chaotic maps. \emph{J.\ Stat.\ Mech.\ Theory Exp.} \textbf{2011}, \emph{2011}, P04021.
\url{https://doi.org/10.1088/1742-5468/2011/04/P04021}

\bibitem{CR12}
Colangeli, M.; Rondoni, L. Equilibrium, fluctuation relations and transport for irreversible deterministic
dynamics. \emph{Physica D} \textbf{2012}, \emph{241}, 681. \url{https://doi.org/10.1016/j.physd.2011.12.005}

\bibitem{RTV00}
Rondoni, L.; T\'el, T.; Vollmer, J. Fluctuation theorems for entropy production in open systems.
\emph{Phys.\ Rev.\ E} \textbf{2000}, \emph{61}, R4679. \url{https://doi.org/10.1103/PhysRevE.61.R4679}

\bibitem{CGLS14b}
Cristadoro, G.; Gilbert, T.; Lenci, M.; Sanders, D.P. Measuring logarithmic corrections to normal diffusion in
infinite-horizon billiards. \emph{Phys.\ Rev.\ E} \textbf{2014}, \emph{90}, 022106.
\url{https://doi.org/10.1103/PhysRevE.90.022106}

\bibitem{MT16}
Marklof, J.; T\'oth, B. Superdiffusion in the periodic Lorentz gas. \emph{Commun.\ Math.\ Phys.} \textbf{2016}, \emph{347},
933. \url{https://doi.org/10.1007/s00220-016-2578-y}

\bibitem{ZPFDB18}
Zarfaty, L.; Peletskyi, A.; Fouxon, I.; Denisov, S.; Barkai, E. Dispersion of particles in an infinite-horizon Lorentz gas.
\emph{Phys.\ Rev.\ E} \textbf{2018}, \emph{98}, 010101(R). \url{https://doi.org/10.1103/PhysRevE.98.010101}

\bibitem{B92}
Bleher, P.M. Statistical properties of two-dimensional periodic Lorentz gas with infinite horizon. \emph{J.\ Stat.\ Phys.}
\textbf{1992}, \emph{66}, 315. \url{https://doi.org/10.1007/BF01060071}

\bibitem{D12}
Dettmann, C.P. New horizons in multidimensional diffusion: The Lorentz gas and the Riemann hypothesis. \emph{J.\ Stat.\
Phys.} \textbf{2012}, \emph{146}, 181. \url{https://doi.org/10.1007/s10955-011-0397-2}

\end{thebibliography}
\end{document}